\documentclass{emulateapj}
\usepackage{colordvi}

\newcommand{\msun}{$M_{\odot}$}

\newcommand{\feh}{[Fe/H]}
\def\f0{$F_0$}
\newcommand\tree[4]{($#1_{#2}$/$1_{#3}$/$1_{#4}$)}
\input epsf

\begin{document}
\title{Chemical Evolution in Hierarchical Models Of Cosmic Structure I:
Constraints on the Early Stellar Initial Mass Function}
\author{JASON TUMLINSON}\affil{Department of Astronomy and Astrophysics,
University of Chicago, 5640 S. Ellis Ave., Chicago, IL 60637}
\begin{abstract}
I present a new Galactic chemical evolution model motivated by and
grounded in the hierarchical theory of galaxy formation, as expressed
by a halo merger history of the Galaxy. This model accurately reproduces
the ``metallicity distribution function'' (MDF) for Population II stars
residing today in the Galactic halo. Model MDFs are calculated for a
fiducial Galaxy formation scenario and a range of assumptions about the
astrophysics of star formation and chemical enrichment at early times.
The observed MDF and the apparent absence of true Population III stars
from the halo strongly imply that there is some critical metallicity,
$Z_{crit} \lesssim 10^{-4} Z_{\odot}$, below which low-mass star formation
is inhibited, and perhaps impossible.  The observed constraints from the
halo MDF, relative metal abundances from extremely metal-poor Galactic
halo stars, and the ionizing photon budget needed to reionize the IGM
together imply a stellar initial mass function (IMF) that is peaked in
the range of massive stars that experience core-collapse supernovae,
with mean mass $\langle M \rangle = 8 - 42$ \msun.  This mass range is
similar to the masses predicted by models of primordial star formation
that account for formation feedback.  A set of five plausible IMF cases
is presented, ranging from broadly peaked with mean mass $\sim 20$
\msun, to narrowly peaked at mean mass $\sim 70$ \msun.  These IMF cases
cannot be distinguished formally by the available constraints, but the
models with lower characteristic mass produce overall better fits to
the available data. The model also implies that metal-poor halo stars
below [Fe/H] $\lesssim -3$ had only 1 - 10 metal-free stars as their
supernova precursors, such that the relative abundances in these halo
stars exhibit IMF-weighted averages over the intrinsic yields of the
first supernovae. This paper is the first part of a long term project to
connect the high-redshift {\em in situ} indicators of early star formation
with the low-$z$, old remnants of the first stars.  \end{abstract}


\section{Introduction}

Interest in the first stars is mounting as the frontiers of formerly
disparate fields of observing and theory converge at the end of the
cosmological ``Dark Ages''. Despite theoretical progress and rapid
advances in the discovery of high-redshift galaxies, many questions
remain open: What is the first-stars IMF? How long did the epoch of
metal-free stars last? Did they contribute much to reionization? Do
their remnants include compact objects, and if so, where are they?
Theorists approach these problems by running the clock forward in
simulations of individual stars and primordial galaxies (Tumlinson \&
Shull 2000; Bromm \& Larson 2004). Observers work backward in time, where
the frontier marked by the most distant known galaxy is steadily advancing
thanks to the {\em Hubble Space Telescope} (HST) and large ground-based
telescopes and is expected to advance rapidly with the launch of {\em
James Webb Space Telescope} (JWST).  The highly interdisciplinary ``first
stars'' field is just beginning to confront observations, in the form of
reionization tests with the Gunn-Peterson effect and the {\em Wilkinson
Microwave Anisotropy Probe} ({\it WMAP}) and chemical abundances from
extremely metal-poor Galactic stars. These indicators have yielded some
indirect constraints on the nature of the first stars.  In the next few
years theory will come into closer contact with high-redshift galaxy
surveys, and additional insight will come from the growing numbers of
old stars in the Milky Way that reveal the detailed metal yields of
early supernovae. But neither high-$z$ galaxies nor the ``second stars''
will tell the whole story on their own. The most promising theoretical
approach is to connect unique high-$z$ signatures of the first stars
{\em in situ} and the Galactic ``fossil record'' preserved in the second
stellar generation.

Following this approach, this paper advances three main goals:
\begin{itemize}
\item[1.]  A formalism for connecting Galactic chemical evolution
         to the theory of galaxy formation, so that different
         observational data can be integrated together.
\item[3.]  An empirical method for assessing the duration and intensity
           of metal-free star formation, to complement and anticipate
           direct observations, and
\item[2.] A framework for relating high-redshift galaxy surveys and spectra 
            to Galactic chemical evolution. 
\end{itemize}

This study extends the approach used by, e.g.~Tumlinson, Venkatesan, \&
Shull (2004, hereafter TVS04), who attempted to constrain the first-stars
IMF by combining the {\it WMAP} constraint on early reionization and
the observed metal abundances in metal-poor Population II halo stars.
That study argued that the joint constraints were best fitted by a first
generation with $10 - 100$ \msun, and not the very massive stars (VMSs;
$M \gtrsim 140$ \msun) motivated by some simulations. TVS04 demonstrates
the potential of relating signatures of reionization at high redshift
with metal-poor stars in the Galaxy, an approach that has been called
``near-field cosmology'' (Freeman \& Bland-Hawthorn 2002).  This
introduction provide some additional motivation in terms of recent
theoretical and observational advances.

{\em Goal 1: Connecting Early Chemical Enrichment to Distant Galaxies.}
The first and most fundamental project will complete the theoretical
framework needed to connect high- and low-$z$ observations. The major
missing piece is a formalism that will follow early chemical enrichment
in the proper context of galaxy formation.  This framework must relate
integrated galaxy properties to the detailed metallicity history of their
stellar populations. A hierarchical, $\Lambda$CDM-based stochastic treatment
is needed because early chemical enrichment may proceed unevenly.
The earliest pregalactic dark-matter halos are thought to be too small,
with $\lesssim 10^3 - 10^4$ \msun\ in stars, to contain a representative
sample of stars from the IMF, especially if they are all massive stars.
The Press-Schechter theory underlying most semi-analytic models of
reionization (e.g., Haiman \& Loeb 1997; Venkatesan, Tumlinson, \& Shull
2003) and the chemical enrichment of the IGM (Scannapieco et al. 2003)
can test a wide range of uncertain astrophysical assumptions, but by
itself results in an unsuitably deterministic and uniform treatment of
early chemical enrichment because it generally treats alike all halos
of equal mass. On the other hand, numerical simulations can follow
complicated gas physics in three dimensions, but often lack the mass
resolution and speed that is required to realize a wide range of
physical assumptions. 

A better model would combine these approaches, and follow the stochastic
history of stellar populations in semi-analytic models.  Such a stochastic
semi-analytic treatment (e.g., halo-merger trees; Somerville \& Primack
1999) is widely used to study galaxy formation and generally succeeds at
reproducing galaxy properties.  In broad outline, the halo-merger model
will calculate the collapse of baryons into protogalaxies, hierarchical
mergers into galaxies, and star formation histories over a range of final
($z$ = 0) galaxy masses, given the astrophysical parameters. Stellar
evolution and atmosphere models (Tumlinson, Shull, \& Venkatesan 2003;
hereafter TSV03) and detailed metal yields (Heger \& Woosley 2002,
hereafter HW02; Umeda \& Nomoto 2005) will specify the ionizing photon
output and the global chemical evolution of the stars and gas. The
resulting model will be capable of predicting both (1) color and
metallicity diagnostics and evolution for high-z galaxies for comparison
to {\em HST} and {\em JWST} data at $z > 6$, and (2) a model distribution
of metal-poor Galactic stars for comparison to the existing samples
(Cayrel et al. 2004; Cohen et al. 2004) and eventually to the thousands
of metal-poor halo stars expected from large future surveys (SEGUE,
WFMOS). Thus, a single model can test astrophysical assumptions for
metallicity evolution and IMF of the first stars against both the
high- and low-$z$ data. This approach is new; the necessary level
of detail in abundances cannot be captured by numerical simulations,
and the galaxy-formation context is missing from conventional chemical
evolution models.

{\em Goal 2: The End of the First Stars Epoch: Transition to Normal Star
Formation.} By connecting early galaxy formation to Galactic chemical
evolution, the complete framework will enable a new approach to what is
perhaps the thorniest problem about the first stars - how long did they
last? The final model will constrain the duration of the first-stars
epoch with the number counts and element abundances in the metal-poor
stars, by deriving a formation history of metal-free stars consistent
with Galactic chemical evolution.

Recent theoretical work has focused on {\it a priori} estimates of the
``critical metallicity'', $Z_{\rm cr} \sim 10^{-5} - 10^{-3} Z_{\odot}$
(Schneider et al. 2002; Bromm \& Loeb 2003; Santoro \& Shull 2005), above
which gas can cool and fragment to $\sim 1$ M$_{\odot}$.  Below $Z_{cr}$
a ``top-heavy'' IMF is expected, owing to the effective absence of
metal-line and dust cooling (Bromm, Coppi, \& Larson 2001; Abel,
Bryan, \& Norman 2002). The time required for a galaxy to achieve this
metallicity, by itself or by enrichment from its neighbors, is believed
to be 100 - 200 Myr (TVS04), but it is difficult to calculate this time
from first principles because a number of poorly-understood processes
(e.g., mixing in the interstellar medium, transport of metals across
the IGM) influence the result. The proposed galaxy formation model
will follow these processes to provide an independent estimate by
asking instead when the ``second stars'' - the Galactic Population II
stars - were born.  The observed abundances of key elements produced
by zero-metallicity stars may be just such a ``clock''.  Ultimately,
halo merger histories calibrated by high-$z$ observations can assign a
timeframe to the metal-poor stars based on their levels of enrichment
([Fe/H]) and relative abundances in the context of the global model.
The approach of relating all available observational data is motivated
by this promise of providing powerful logical connections between all
constraints, direct and indirect, present and future.

{\em Goal 3: Explain the Signatures of Early Reionization and
High-redshift Galaxies.} As high-$z$ galaxy searches expand to encompass
larger, deeper samples, they will eventually enter the transition period
between metal-free and metal-enriched star formation and into the epoch
of reionization at $z \gtrsim 6$. There are already indications from the
cosmic microwave background (CMB) that reionization began earlier and
was more extended in time than had been expected on theoretical grounds
alone. This early reionization can be explained by a variety of models,
but the most commonly cited explanation is that early star formation is
unusually efficient at converting stellar mass into ionizing photons
(Cen 2003; Haiman \& Holder 2003; Ciardi et al. 2003; TVS04). Thus the
IMF of early star formation is intimately connected with reionization,
and it is worth asking whether the populations that accomplished
reionization have left any detectable information in the history of
Galactic chemical evolution.  TVS04 argued for such a connection, and in
favor of mutual constraints on the primordial IMF from reionization and
chemical evolution.  In addition to developing the basic formalism for
goals 1 and 2, this paper carries this effort further toward quantitative
constraints.

The need for robust metallicity indicators for early galaxies will soon
become acute. The unique signatures of truly metal-free populations are
expected to be the high Ly$\alpha$ equivalent width (which may have been
detected already at $z \sim 5$; Malhotra \& Rhoads 2002) and strong He II
$\lambda$1640 nebular emission (Tumlinson, Giroux, \& Shull 2001); these
sensitive indicators of stellar IMF from $10 - 500$ \msun\ are also unique
to $Z = 10^{-4} Z_{\odot}$ (Schaerer 2002; TSV03). As shown by TSV03,
the broadband colors of metal-free stellar clusters are easily confused
with more metal-rich populations, so Ly$\alpha$ and He II are likely to
prove essential to finding the true ``first stars''. By coupling early
chemical enrichment to evolving stellar spectra, the final model will
provide detailed connections between early chemical evolution preserved
in the Galactic halo and direct indicators at high redshift.

These three projects are all long-term goals of a final model only the
beginning of steps of which are reported here, mainly related to Goal
1. This paper focuses on integrating Galactic chemical evolution into
the hierarchical theory of galaxy formation.  For now, I concentrate my
efforts on laying out the model in detail and comparing its results for
the Galaxy against detailed constraints from metal-poor stars and the
requirements of reionization.  As explained below, the initial focus is on
deriving empirical constraints on the first-stars IMF, as a complement to
purely theoretical approaches that are being pursued by many groups (Bromm
et al. 2001; Abel et al. 2002; Omukai et al. 2005; Tan \& McKee 2004).

Theory is conflicted on whether the first stars are exclusively very
massive (I adopt here the formal definition $M \geq 140$ \msun\ for VMSs -
see \S~3.1.3).  Abel, Bryan, \& Norman (2002) first implemented rigorous
H$_2$ cooling in a cosmological hydrodynamical simulation and found
that the first star formed in their simulated volume was surrounded
by several hundred solar masses of material that could have accreted,
leading to the suggestion of stars with only $M \gg 100$ \msun\ in the
first generation.  Recent studies (Omukai \& Palla 2002; Tan \& McKee
2002), taking the Abel et al. result as a starting point for semi-analytic
models of primordial star formation, indicate that primordial star
formation may be complicated by feedback on the accreting material.
Omukai \& Palla (2002) identified a critical accretion rate ($\dot{M}$
$ = 4 \times 10^{-3}$~\msun\ yr$^{-1}$) above which laminar accretion
onto a spherically-symmetric protostar exceeds the Eddington limit, when
the star has $M = 100 - 300$ \msun.  Models by Bromm \& Loeb (2004)
found a conservative upper limit of 500 \msun\ by setting the time
available to accretion to be the stellar lifetime ($\sim 3 \times 10^6$
yr) but ignoring feedback. Tan \& McKee (2002) consider the effects of
disk accretion geometry, rotation, and radiation feedback in limiting
the mass of primordial stars. They find that these feedback mechanisms
are likely to operate at $M = 30 - 100$~\msun, perhaps limiting the
masses of metal-free stars to this range.  However, they later found
(Tan \& McKee 2004) that none of the feedback mechanisms considered can
halt the accretion before the star achieves $\sim 30$ \msun.  Omukai et
al. (2005) considered the effects of metals on the fragmentation of
low-metallicity clouds, and suggest that a purely top-heavy IMF at
$Z = 0$ switches to a bimodal IMF at finite metallicities $Z \lesssim
10^{-4}$ before settling to a more normal power-law function near $Z \sim
10^{-2}$. These important results indicate that the mass limits of the
primordial IMF may be quite different from today. This intriguing yet
unsettled guidance from state-of-the-art theory motivates this empirical
approach to constraining the IMF of the first stars.

The IMF, as it is classically understood in the Galaxy today, is derived
from counting stars in Galactic stellar clusters. In some cases a
single cluster is massive and young enough to sample the full IMF
(see Figer 2005 for a state-of-the-art case). This technique is still
feasible in regions of the local universe close enough to allow imaging
of resolved stars, such as the Magellanic Clouds.  Beyond this, the IMF is
an abstraction that must be understood in a model-dependent fashion using
indirect indicators such as galaxy colors and emission from \ion{H}{2}
regions. In these cases it may still be sensible to speak of the IMF
as representing the spectrum of masses in a single, coeval cluster of
stars. In the early universe, within small dark-matter halos that may
have only $10^5 - 10^7$ \msun\ of gas and 100 - 1000 \msun\ in stars, the
IMF takes on a third distinct meaning. Here it is the parent probability
distribution from which stellar masses are drawn stochastically based
on their (indeterminate) initial conditions for star formation. As the
initial conditions for star formation may vary from place to place,
the parent probability distribution may also vary. However, theoretical
efforts have identified the gas metallicity as the critical factor
influencing the cooling, collapse, and fragmentation properties of
low- or zero-metallicity gas (Bromm et al. 2002; Omukai et al. 2005),
it is sensible to speak of a single probability distribution for
zero-metallicity gas. Thus, the IMF is taken here to be the probability
distribution of masses for stars formed from metal-free gas, wherever
it exists. 

\begin{deluxetable}{ccl} 
\tablecolumns{3} 
\tablewidth{0pc} 
\tablecaption{Model Parameters and Their Fiducial Values}
\tablehead{Parameter & Value & Description }
\startdata
\cutinhead{Cosmology}
$\Omega _M$ & 0.27\tablenotemark{a} & Cosmic matter density \\
$\Omega_\Lambda$ & 0.73\tablenotemark{a}  & Cosmological constant \\
$h$ & 0.71\tablenotemark{a}  & Hubble parameter \\
$\Omega _b$ & 0.044\tablenotemark{a}  & Baryon density \\
$\sigma _8$ & 0.84\tablenotemark{a}  & Power spectrum normalization \\
$\delta _{c,0}$ & 1.686 & Critical overdensity for collapse \\
\cutinhead{Chemical Evolution}
$M_{\rm gal}$  & $5 \times 10^{11}$ \msun\ & Total Galaxy mass \\
$\tau_{disk}$ & $0.8 - 3 \times 10^8$ yr & Disk formation timescale \\
$z_{disk}$    & 8 - 12 & Halo cutoff redshift \\ 
$T_{vir}^{\rm min}$ & 10$^3$ - 10$^4$ K & Minimum virial temperature \\
$M_l$         & $10^8$ \msun\ & Minimum mass resolution \\ 
$\epsilon_{*}$ & $2 \times 10^{-10}$ & Star formation parameter \\
$m_{\rm Fe}$ & 0.07 \msun\ & Standard Type II iron yield \\
$Z_{\rm cr}$ & $10^{-4}$ & Critical metallicity \\
$M_{dil}^0$   & $10^6$ \msun & Interstellar dilution mass \\
$t_{dil}^0$   & $10^7$ yr & Interstellar mixing timescale \\
\cutinhead{Intergalactic Medium}
$f_{esc}^{Z}$ & 0.05 & Mass fraction of ejected metals \\
$M_{dil}^{IGM}$ &  $5.0 \times 10^8$ \msun\ & Final IGM metal dilution mass \\
$t_{dil}^{IGM}$ &  $5.0 \times 10^9$ yr & IGM metal dilution timescale
\enddata
\label{galtable}
\tablenotetext{a}{Spergel et al. (2003)}
\end{deluxetable}

The terminology in this paper maintains a distinction not usually seen
in the theoretical literature. The designations ``Population II''
and ``Population III'' are historical observational definitions referring
to detected and hypothesized populations residing the Galaxy today
(Baade 1944). This definition is maintained here, and these terms
are used to denote only Galactic stars. The terms ``metal-free''
or ``zero-metallicity'' will refer to early stellar populations more
generally by including those stars born early that were too short-lived
to be around today. By this definition, a true Population III star is
a (presumably low-mass) star with no metals that resides today in the
Galactic halo. To date no Population III stars have been found.

I emphasize here that what follows is not a rigorous model for the
formation of the Galaxy, intended to represent faithfully the actual mass
assembly history, and in particular the dynamics, of the Galaxy. This
is a model for Galactic chemical evolution following in the tradition
of analytic chemical evolution models but motivated by and grounded
in the hierarchical theory of galaxy formation. The model was created
to take full advantage of present and future observational efforts
to dissect the detailed assembly history of the Galaxy and to try to
understand near-field signatures of early stars. Since the initial
effort has met with some success, this model will, in the future,
be extended to include dynamics and to cover a population of galaxies
at arbitrary redshift, to connect it to high-redshift galaxy surveys
(Goals 2 and 3). These extensions represent another leap in complexity,
detail, and computing time, and so will be deferred to later papers.
Until then, the reader must keep in mind the motivation and limitations
of the model in assessing the conclusions.

The paper proceeds as follows. Section 2 describes the merger-tree
structure formation model in detail. Section 3 describes the chemical
evolution model including the assumptions and their motivations. Section
4 describes observational information that will be used to constrain the
primordial IMF. The major results on the IMF are described in Section
5. Section 6 describes some additional results of the model that do
not involve the IMF directly, including information about the epoch of
the first stars.  Section 7 discusses these results in the context of
other theoretical results and present and future observational efforts to
address these questions. Section 8 summarizes the key results and provides
some additional comments on the future of studying the first stars.

\begin{figure*}
\centerline{\epsfxsize=\hsize{\epsfbox{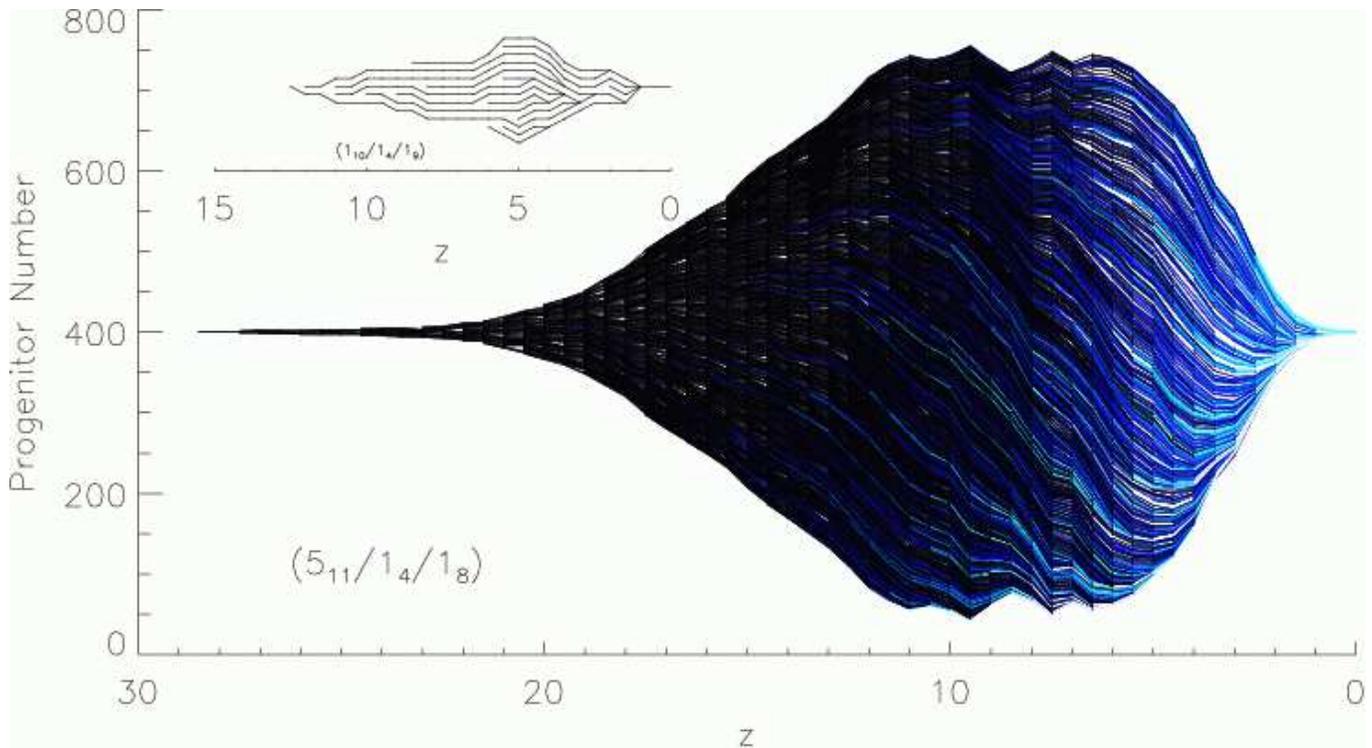}}}
\figcaption{Examples of two halo merger trees. At $z = 0$, the larger
final halo has $M_h = 5 \times 10^{11}$ \msun. The tree was calculated with
virial temperature resolution $T_{vir}^{min} = 10^4$ K and mass resolution
$M_l = 10^8$ \msun. The number of active progenitor halos peaks at $z \sim
6 - 10$, beyond which halos begin to drop below the finite virial temperature
resolution, become ``leaf'' halos and so disappear from the tree. The
color scale for the main tree expresses the halo mass increasing from
black to light blue. The single halo at $z = 23$ is the first-born
progenitor of the Galaxy's most massive immediate predecessor. The
inset shows a smaller tree for illustration of typical branching
behavior. \label{treefig}} \end{figure*}

\section{A Hierarchical Model for the Growth of Structure}
\label{epssection}

The chemical evolution model consists of two major components. First,
a model of hierarchical structure formation is used to decompose the
present-day Galactic dark-matter halo into its constituent pieces,
working backward in time. Once this process is complete the second
component uses this mass assembly history to calculate histories for
stellar populations and chemical enrichment, working forward in time. The
structure formation component is described in this section, and the
chemical enrichment calculations are described in \S~\ref{chemsection}.

This section gives a brief introduction to the basic theory underlying the
structure-formation model; more detail about the specific implementation
is given in the Appendix.  This model relies on a framework for following
the hierarchical growth and mergers of dark matter halos over time
that has been fully developed in the literature over the last three
decades. The model primarily follows the method and notation of Lacey
\& Cole (1993; hereafter LC93), who developed the ``extended
Press-Schechter'' formalism for calculating halo merger probabilities and
rates from the more {\em ad hoc} treatment of Press \& Schechter (1974).
For a thorough introduction and rigorous derivations of these relations,
see LC93 or Liddle \& Lyth (2000).

The extended Press-Schechter treatment of structure formation begins
with a power spectrum of Fourier modes, $P(k)$, that describes the
probability distribution of small ($\delta \equiv \rho/\langle \rho
\rangle \ll 1$) density fluctuations in the early universe. From early
times these fluctuations grow linearly according to the linear growth
function, $\delta(x,t) = \delta(x,t_0)D(t)/D(t_0)$, until $\delta \sim
1$, when the fluctuation is said to have entered the nonlinear regime,
``turned around'' and virialized. The overdensity of turnaround
corresponds to a critical linear overdensity, $\delta_c$, that can be
calculated by considering the evolution of an isolated spherical region
(see Section 11.1 of Liddle \& Lyth 2000). This linear density contrast
is only weakly dependent on cosmological parameters; here $\delta_c =
1.686$. This critical density separates regions in the evolving linear
field that have collapsed from those that have not.

The number density of collapsed halos is equivalent to number of
density fluctuations that have achieved the critical linear density
by the redshift of interest. The density field is first smoothed over
some mass scale $M$ to derive the variance of the density field as a
function of smoothing mass, $S(M)$, using a spherical top-hat filter
and performing the calculation in Fourier $k$-space:
\begin{equation}
S(M) \equiv \sigma^2(M) = \frac{1}{2\pi^2} \int^{\infty}_{0} P(k) \bar{W}^2(kR)k^2 dk, 
\end{equation} where $\bar{W}(kR)$ is the real-space top-hat
window function. Further details of this calculation are given in the
Appendix. From now on, I use the local (smoothed) variance as a function
of mass, $S(M)$, to describe the density field.  Generally, $S(M)$
decreases as $M$ increases and small wavelength density fluctuations are
smoothed out.  From here, the calculation will be considered in terms of
a single mass element or test particle as the smoothing mass is varied,
rather than the variations from place to place.  At a given location,
the variance $S(M)$ executes a random walk as the smoothing mass is
varied, occasionally crossing the critical barrier of $\delta_c$, where
it is said to have collapsed.  To avoid the so-called ``cloud-in-cell''
problem (LC93), the largest mass at which a given trajectory crosses
this barrier is taken as the mass of the collapsed halo inside which
the test particle resides.  The number density of regions that lie above
$\delta_c$ is described by a diffusion equation with an absorbing boundary
condition at $\delta_c$, leading to an expression for the number density
of collapsed objects as a function of mass, the Press-Schechter formula:
\begin{equation}
ndM = \sqrt{\frac{2}{\pi}} \frac{\bar{\rho}}{M} \frac{\delta _c
      (z)}{\sigma ^2 (M)} {\rm exp} \left[- \frac{\delta ^2 _c (z )}{2 \sigma ^2
      (M)}\right] dM
\end{equation} where $\bar{\rho}$ is the cosmic mean matter density. 
Further manipulation of this formula (LC93) gives the so-called ``conditional
mass function'', or the fraction of possible trajectories inside collapsed
mass $M_1$ at redshift $z_1$, that are also in halos with larger mass $M_0$
at lower redshift $z_0$. This relation is given by LC93:
\begin{equation}
P(\Delta S, \Delta \omega) d\Delta S = \frac{1}{\sqrt{2\pi}}
 \frac{\Delta\omega}{(\Delta S)^{3/2}}
 \exp{\left[-\frac{(\Delta \omega)^2}{2 \Delta S}\right]} d\Delta S
\label{eq-ps}
\end{equation}
With a change of variables
\begin{equation}
x = \Delta \omega / \sqrt{\Delta S},
\label{eq-varchange}
\end{equation}
this probability distribution assumes the form of a Gaussian with
unit variance.  Given a known redshift timestep ($\Delta z \propto
\Delta \omega$), this formula describes the probability distribution of
``child'' halos which compose the larger halo at lower redshift. This
formula can be used to stochastically decompose a large low-$z$
halo progressively into smaller and smaller pieces in a halo merger tree.

The method for constructing trees follows the ``N-fold merger with
accretion'' method devised by Somerville \& Kollatt (1999).  The
construction of a merger tree begins with the choice of a final halo mass,
$M_{h}$, at $z = 0$. For a model Milky Way, I choose $M_{h} = 5 \times
10^{11}$ \msun\ (the final results will be shown to be insensitive to
this choice). The merger tree is specified by $M_{h}$ and
the assumption of mass resolution and/or minimum virial temperature.
At each redshift, the mass resolution is the smaller of $M_l$ and
$M(T_{vir}^{min})$.  Beginning with the final halo at $z = 0$, I
first derive a redshift interval over which to evolve the halo (the
``timestep'' equivalent to $\Delta \omega$, where $\omega (z) = \delta_c
(z)$, see Appendix).  This can be set to either a fixed value (usually
$\Delta z = 0.5$) or allowed to be redshift and/or mass dependent. This
choice slightly affects the overall size of the tree. 

Once the timestep is determined, the ``parent'' halo is disintegrated into
its ``child'' halos according to the ``N-fold with accretion'' scheme
developed by SK99.  A random variant is drawn from Equation~\ref{eq-ps}
after the change of variables and converted into child mass using
equations Equations 1, 3, and 4.  If this mass is greater than the smaller
of $M(T_{vir}^{min})$ and $M_{l}$, this halo is termed a ``child'' and
its mass is subtracted from the reservoir. If this new halo mass falls
below the mass cutoff, the random variant is drawn again. This procedure
is repeated until the remaining mass of the parent is less than $M_{l}$
or $M(T_{vir}^{min})$, at which point the parent halo has been divided
into typically 2 - 5 progenitor halos and a parcel of ``accreted''
material. Halos that lie below $M_l$ and/or $T_{vir}^0$ at their redshift
are termed ``leaf'' halos. The mass limit $M_l$ is imposed to inhibit a
large number of very low-mass progenitor halos at low redshift. A mass
integral over all leaf halos and accreted mass equals the total mass of
the final galaxy halo, $M_h$.

In the following discussion individual trees are labeled with their
final halo mass $M_h$, minimum virial temperature $T_{vir}^{min}$,
and absolute mass resolution $M_l$ as ($M_h$, $T_{vir}^{min}$,
$M_l$) using exponential notation such that $1_8 = 1 \times 10^8$.
Thus a \tree{5}{11}{3}{8} tree models a dark matter halo with mass $5
\times 10^{11}$ \msun\ at $z = 0$, decomposed into progenitor halos with
minimum virial temperature $10^3$ K or minimum mass $M_l$, whichever is
lower at a given redshift.

To ensure that the method produces a distribution of halo masses
that matches the original Press-Schechter function for the number
density of halos, I create a ``universe'' of trees by drawing a large
number (typically 10000) halos from the analytic EPS distribution at $z =
0$. Merger histories are then calculated for these halos and combined
into a single distribution for comparison to the EPS function at any
redshift (Equation 2). Good agreement is found between the analytic halo mass
spectrum and the numerical results. The numerical results calculated
from the EPS formalism also agree well with the halo number-density and
mass distributions presented by Mo \& White (2002) when their fiducial
cosmological parameters are adopted.

\section{Chemical Evolution in the Hierarchical Context}
\label{chemsection}

This section describes the chemical evolution model, including its
relationship to the merger trees, the methods for calculating chemical
evolution, and their assumptions. For halos in the merger tree, there
are in essence only two processes that must be followed. These are halo
evolution in isolation (\S~\ref{evolvesection}) and halo
mergers (\S~\ref{mergersection}). In addition, the transport of metals
from virialized halos into new halos is followed in terms of a model IGM,
which is described in \S~3.3.

To calculate a chemical enrichment history for a merger tree, the driver
starts with the earliest ``child'' halo ($z = 23$ in Figure~\ref{treefig})
and evolves it forward in time until a merger is reached. If all the
child halos involved in that merger have had their evolution calculated,
the merger occurs. If the child halos are not yet all calculated, the
driver steps back to higher redshift until it reaches a leaf halo and
uses that as a starting point for working forward until it reaches
the potential merger where it stopped before. If all components of
the potential merger are now calculated, the merger occurs. If not,
the driver steps back, and so on. In this way the driver calculates the
chemical enrichment history of all individual halos and merges them,
such that the final $z = 0$ halo is the last to be merged.

\subsection{Chemical Evolution of Isolated Halos}
\label{evolvesection}

Between mergers, a halo in isolation evolves forward in time according
to equations describing its stellar populations and gas mass and
metallicity. Inside an isolated halo, these equations simply describe
the rate at which gas is formed into stars, the IMF, the rates at which
mass is returned to the ISM (in the form of ejected gas and metals),
and the properties of mixing in the ISM. This evolution is managed by
a driver which takes 10 - 50 uniform timesteps between the initial and
final redshifts for the halo.  I describe these methods here in detail.

\subsubsection{Star Formation Law}

Gas residing in gravitationally bound dark matter halos is assumed
to form stars at a fixed mass efficiency per year, $\epsilon_{*}$.
The total mass in stars formed in a given time interval $\Delta t$
is given by $M_* = M_{gas} \times \epsilon_{*} \times \Delta t$.
The fiducial rate is fixed at $\epsilon_{*} = 2.0 \times 10^{-10}$ for
all halos to approximate the total gas mass ($5 \times 10^9$ \msun) and
stellar mass in the Milky Way for the final halo of a \tree{5}{11}{3}{8}
tree.  Star formation is allowed to proceed, and is followed explicitly,
only in those halos which have $M \geq M(T_{vir}^{min})$ at their
redshift. Child halos with less than this mass are considered to be
diffuse ``accreted'' gas, which is placed into the halo in uniform
parcels in each timestep during single halo evolution.

\subsubsection{Disk/Halo Mass Segregation}
\label{diskhalosection}

If all star formation down to $z = 0$ is assumed to take place in the
Galactic halo, and to deposit its stars and ejected metals there, then
the final model halo possesses many stars which properly belong in the
Galactic disk and/or bulge. The semi-analytic galaxy-formation models on
which this model is based (e.g., Somerville \& Primack 1999) typically
track the growth of a disk explicitly based on the angular momentum of
accreted gas parcels, but this method would add too much complication and
parametric freedom to this simple model. To reproduce the Galactic halo
metallicity distribution function (MDF), it is necessary to introduce a
simple prescription for when gas has settled into the disk. Two approaches
to this problem are adopted here.

First, a single parameter, $\tau_{disk}$, can control the time after
which a parcel of gas in the dark-matter halo has settled into the
disk. Each parcel of gas in the tree, whether acquired by accretion or
merger, has a time associated with it that measures the interval since
it was virialized or accreted. When this time has reached $\tau_{disk}$,
the parcel is moved into the disk and no longer forms halo stars. This
parameter is the most sensitive control on the location of the peak for
the halo metallicity distribution function (see \S~4). For $T_{vir}^{min}
= 10^4$ and $10^3$ K, $\tau_{disk} = 0.8 - 3.0 \times 10^8$ yr gives
the best fit to the peak. With this value, halos convert $\tau_{disk}
\times \epsilon_* \sim 2-6$\% of their gas mass into stars before the
gas enters the disk. This approach effectively imposes a fixed timescale
{\it per halo} such that the cutoff occurs over a range of redshifts
reflecting the varying birth times of halos. This approach is not limited
to representing disk/halo segregation only - it can also stand in for the
timescale after which gas no longer forms star because of SNe feedback
or some other mechanism. 

An alternative approach is to impose a fixed redshift after which
all halo star formation ceases.  This fixed redshift cutoff can also
reproduce the basic peaked shape of the MDF (see \S~4.1), for a cutoff
fixed at $z_{disk} = 8 - 12$ depending on $T_{vir}^{min}$.  The disk
timescale method produces a slightly better overall fit to the halo
MDF than does $z_{disk}$. In some cases when both are included the
overall fit is improved over the best fit if only $\tau_{disk}$ is used,
particularly at \feh\ $< -1.0$ (where the data is uncertain - see RN91
and Section~\ref{empsection}).  Developing a rigorous method for mass
segregation will require accurate treatment of the dynamics of the Galaxy
formation process, which are completely absent from the present model but
which will form a key component of a final model. Because the choice does
not affect the primary constraints on the IMF discussed in \S~4, I adopt
$\tau_{disk} = 0.8 - 3 \times 10^8$ yr in the fiducial model, and proceed.

\begin{figure}
\centerline{\epsfxsize=\hsize{\epsfbox{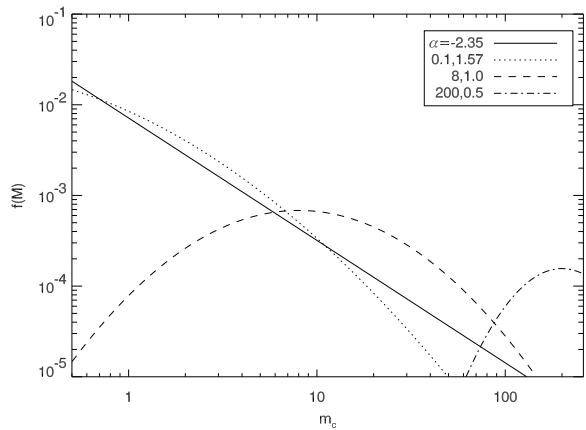}}}
{\figcaption{Examples of four different stellar initial mass functions,
including a power-law IMF with $\alpha = -2.35$ (solid), the Galactic
Miller-Scalo IMF with $m_c = 0.1$, $\sigma = 1.57$ (dotted), a
log-normal IMF with $m_c = 8, \sigma = 1.0$ (dashed), and a ``strong
VMS hypothesis'' IMF with $m_c = 200, \sigma = 1.0$ (dot-dashed), all
normalized to the same total mass.  \label{imf-example-fig}}}\end{figure}

\subsubsection{Stellar Initial Mass Function}

The IMF is the most important ingredient for modeling early
chemical enrichment. The IMF controls both the mass budget of metals
released into the ISM and the number and mass distribution of low-mass
stars that survive to $z = 0$.  The IMF can take two forms: a power
law with high and low mass limits, or a log-normal form with or without
limits.  The power law IMF has the form 
\begin{equation}
\frac{dN}{dm} = m^{-\alpha} 
\end{equation} For a Salpeter or ``standard'' IMF, the power-law slope
is $\alpha = 2.35$ (Salpeter 1955), and at least a lower mass limit
must be assumed to keep the total number of stars finite. This IMF has
the advantage of a simple functional form and general agreement with
observations of the local universe, and the drawback that it is too
simple to describe a potentially wide range of probability distributions
that may obtain in the early universe.  The log-normal IMF (Larson 1973)
takes the form: 
\begin{equation} 
\ln \left( \frac{dN}{d\ln m} \right)= A - \frac{1}{2\sigma ^2}
       \left[ \ln \left( \frac{m}{m_c} \right) \right] ^2
\end{equation} 
where $\sigma$ is the width of the distribution, $m_c$ is the
characteristic mass, and $A$ is an arbitrary normalization.  This IMF
has the advantage of more flexible behavior with only one additional
parameter. This study uses lower and upper mass limits of 0.5 and 260
\msun.  The mean mass for a given pair of log-normal IMF parameters
is slightly higher than the characteristic mass $m_c$ because of the
adopted lower mass limit of 0.5 \msun.

I adopt here the ``VMS hypothesis'', as introduced and defined by
TVS04. Under this definition, ``very massive stars'' (VMSs) are
those with $M \geq 140$ \msun, the smallest mass at which the pair
instability can completely disrupt the star and leave an unusual
nucleosynthetic signature (Heger \& Woosley 2002, hereafter HW02). Stars
with 8 - 140 \msun\ are ``massive stars'' with yield patterns
characterized by Type II supernovae.  This upper limit  for massive
stars is slightly higher than the 100 \msun\ conventionally used in
simple stellar population studies (Starburst 99; Leitherer et al. 1999)
for the Galaxy and the nearby universe, but there are suggestions that
some such stars exist today. Figer (2005) has recently established a
firm upper limit of 150 \msun\ for the Galactic Arches cluster, which
contains at least one star of 130 \msun. Thus 140 \msun\ serves as a
natural dividing line on both theoretical and observational grounds.

There are two different versions of the VMS hypothesis that are
distinguished by their different nucleosynthetic patterns. The ``strong''
VMS hypothesis holds that metal-free stars formed only with $M \gtrsim
140$ \msun, or more technically, that VMS are the only stars in the IMF
that leave a nucleosynthetic signature.  In the ``weak'' VMS hypothesis,
the $8 - 40$ \msun\ progenitors of Type II supernovae form in addition
to VMS. In TVS04, this definition was rigidly constructed and the
two versions were expressed in terms of an IMF that was truncated
at either 140 \msun\ (``strong'') or 10 \msun\ (``weak''). Here the
definition is loosened slightly by dropping the strict truncations
in favor of a log-normal IMF.  Examples of these IMF shapes appear in
Figure~\ref{imf-example-fig}.  The distinction now depends on which mass
range, and therefore which type of supernova, dominates the production of
metals, with ``dominance'' nominally taken to mean one-half the budget
of iron.  These considerations are used below to constrain the early IMF
in detail.

\subsubsection{Critical Metallicity for Normal Star Formation} 

Most of the unique behavior associated with primordial star formation
follows from the unique circumstances of cooling by zero-metallicity
gas.  At $T \lesssim 10^4$~K in the local interstellar medium, cooling
by H and He is negligible and metal-line cooling is dominated by [C~II],
[O~I], and [Si II] fine-structure lines.  Because metal-free gas is
restricted to inefficient cooling by H$_2$ at $T \lesssim 10^4$~K,
cooling rates are reduced, and the balance between pressure support and
gravitational collapse is shifted to higher temperatures.  Primordial
protostellar objects, being unable to cool or fragment, should
therefore be more massive to overcome their elevated gas pressure.
At critical metallicity $Z \gtrsim$ 10$^{-5.5}$ - 10$^{-3.5} Z_{\odot}$,
protostellar clouds are able to cool and fragment more efficiently,
leading to a ``normal'' IMF (Schneider et al.~2002; Bromm \& Loeb 2003).
This metallicity may not be fairly represented by a single value, if the
yields of early SNe are highly variable and the first metal-enriched
star forming regions incorporate only a few SNe in stochastic fashion
(Santoro \& Shull 2005). For simplicity this model assumes that $Z_{crit}$
is expressed as a single \feh\ as a parameter of the model.  There is
a capacity to switch the IMF at a single $Z_{crit}$, which happens at
different times in different halos and so spans a range of redshifts.

\subsubsection{Stellar models and supernova metal yields}

Stellar lifetimes are specified by an analytic fit to the $Z = 0$
H-burning main sequence lifetimes specified in Table 2 of Tumlinson,
Shull, \& Venkatesan (2003).  Type Ia thermonuclear SNe are formally
included with $m_{Fe} = 0.5$ \msun\ but generally enter too late
to affect subsequent star formation in the halo, considering the
timescale for disk/halo segregation.  All massive stars with $M = 8 - 40$
\msun\ experience core-collapse SN at the end of their lifetimes (HW02)
and release 0.07 \msun\ of iron into the parent halo.  Detailed yields
for pair-instability supernovae (PISNe) are taken from HW02.

\subsubsection{Metal dispersal and mixing}

Metal mixing in the interstellar medium is another important factor
in chemical evolution.  The first key assumption about metal dispersal
is that the ``boxes'' that represent dark matter halos are closed and
metals do not escape (except for the small fraction represented by
$f_{esc}^Z$ that escape to the IGM).  This leaves two avenues by which
halos acquire metals. First, metals produced by local SNe are released
into the gas and enrich subsequent generations. Second, when new halos
virialize from the IGM, they bring along gas that may be metal-enriched
(see \S~\ref{igmsection}).  This key assumption of closed boxes has many
effects on chemical evolution and may in fact be violated in some cases
(see discussion of minihalos in Section~\ref{minihalosection}).

Rather than attempt to follow the complicated details of supernova remnant
evolution in a clumpy medium (e.g., Oey 2000), this model accounts for
the cumulative effects of SN enrichment in terms of mass only, and in a
stochastic fashion. There are two basic principles behind the treatment of
mixing: (1) that the mass into which the ejecta of a single supernova are
mixed grows steadily over time before asymptotically approaching a fixed
maximum dilution mass, and (2) that new star formation randomly samples
metals from the parcels of mass enriched by previous generations, allowing
for maximum statistical fluctuations in the metallicity distribution of
new stars. Principle 1 provides a simple mechanism for tracking metal
enrichment that elides many of the complicated details of interstellar
physics. Of course, this approach may omit other useful information,
but it is necessary to make this model tractable and executable in short
times. Principle 2 provides maximum flexibility, in that individual SN
can be treated as completely uncorrelated. Various degrees of correlation
can be introduced to make the model more deterministic and perhaps more
representative of clustered or triggered star formation, while a model
designed to follow clustered or triggered star formation explicitly
could not be tuned for more stochastic behavior.

For each massive or very massive star the time of its SN, $t_{SN}$, is
stored. After this time is reached, the ejected metals are assumed to
have been released into the interstellar medium of the parent
halo, and diluted into a dilution mass at a later time $t$: 
\begin{equation}
M_{dil}(t) = M_{dil}^0 (1 - e^{(t_{SN}-t)/t_{dil}^0}),
\label{dil-eq}
\end{equation}
This ``dilution mass'' is then divided by the halo gas mass to derive
the mass fraction of the total halo mass into which the products of
each supernova have been evenly mixed. Where the dilution mass exceeds
the total gas mass of the parent halo, the mass fraction is truncated
at unity (embodying the assumption of closed boxes). Then a vector of
random variants is drawn with as many elements as there are SNe. Where
these random variants lie within the mass fraction enriched by an SN,
that SN contributes to the total metal budget for the stars formed in that
timestep. For the fiducial models, $M_{dil}^0 = 10^6$ \msun\ and
$t_{dil}^0 = 10^7$ yr. These parameters have been set to express the 
values expected for the growth of supernova remnants in the Galactic 
ISM during the radiative phase (Spitzer 1978). This generally leads to
a monotonic enrichment for massive halos but can yield highly stochastic
results in the first small halos. In the limit of small $t_{dil}^0$ or
large $M_{dil}^0$, the mixing approaches an instantaneous limit that is
useful for exploring some limiting cases.

\subsection{Merging multiple halos}
\label{mergersection}

Halo mergers are pairwise sums over all the progenitor halos and so
simply inherit all their progenitors' stars and gas. Mergers affect the
evolution only indirectly by providing a larger dilution mass for metals
previously released into the gas reservoirs of the merger partners.

\subsection{The Intergalactic Medium}
\label{igmsection} 

\begin{figure*}
%\centerline{\epsfxsize=0.7\hsize{\epsfbox{f3a.eps}}} 
%\centerline{\epsfxsize=0.7\hsize{\epsfbox{f3b.eps}}} 
\plottwo{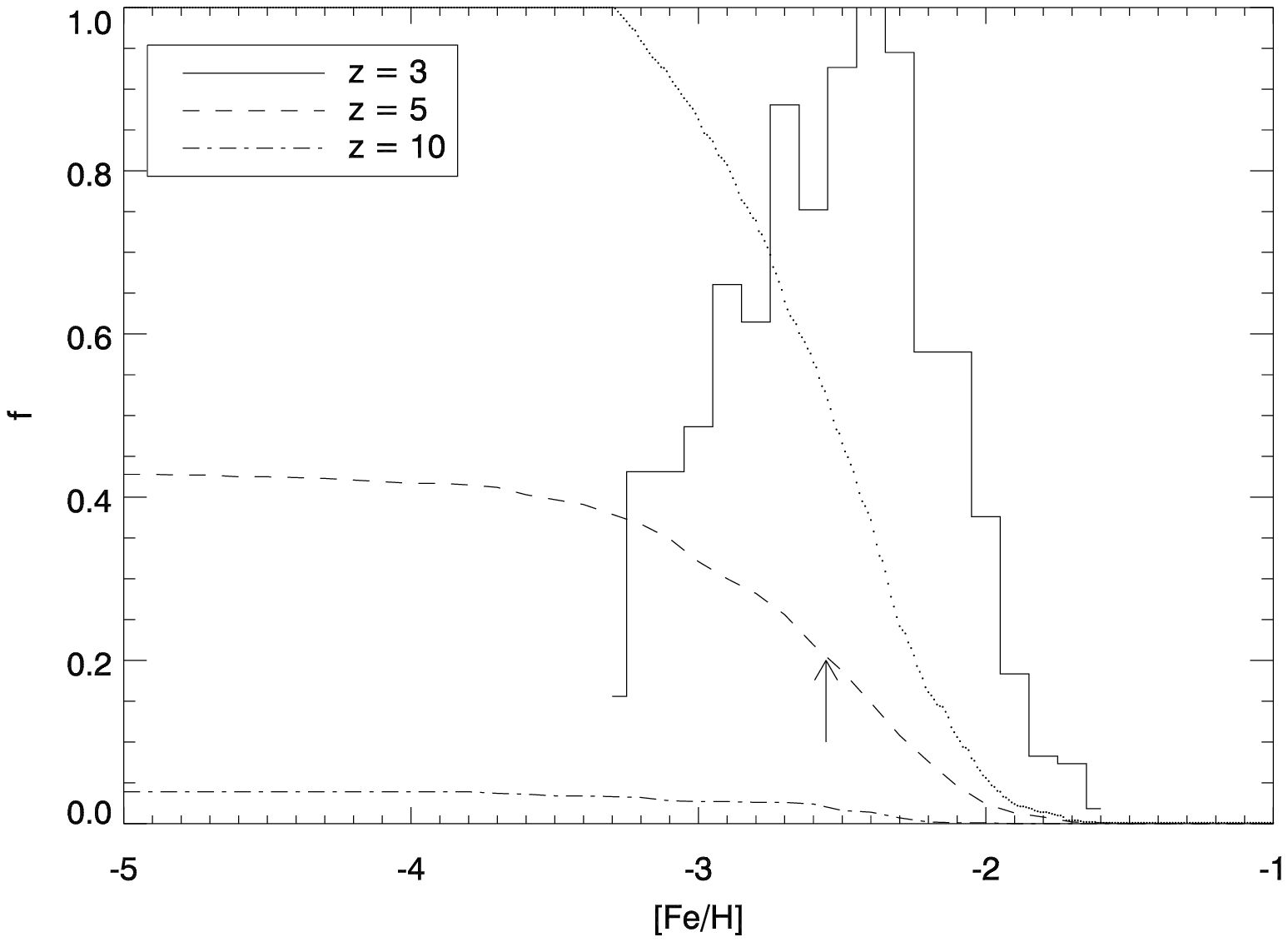}{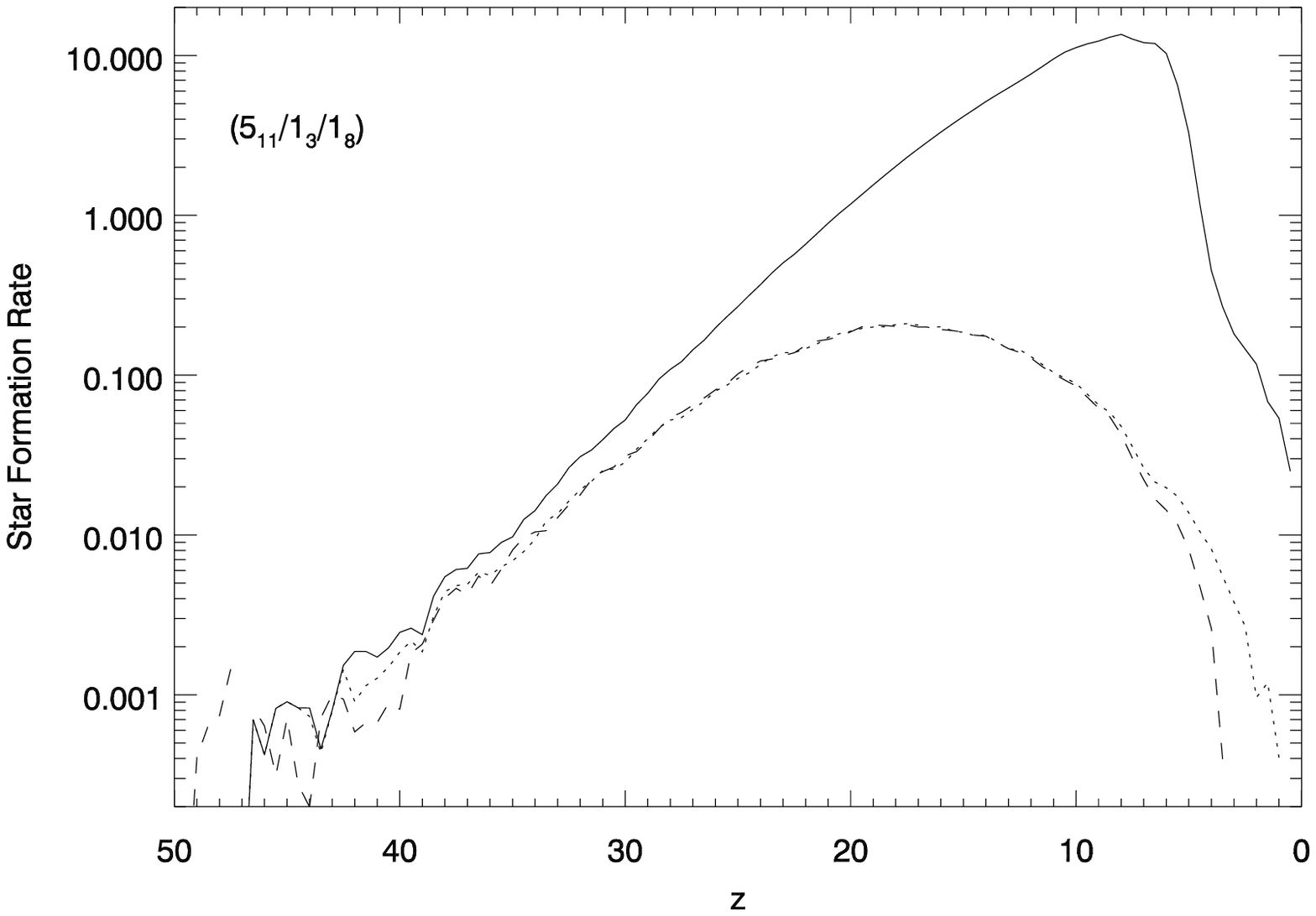} 
\caption{Left panel: The cumulative distributions of IGM metallicity
$f(Z)$ from which new halos are drawn, shown at $z = 3$ (dotted), $z
= 5$ (dashed), and $z = 10$ (dot-dashed) for the fiducial parameters
($f_{esc}^{Z}$, $M_{dil}^0$, $t_{dil}$) = (0.05, $5_8$, $5_9$).  This
model has been tuned to approximate the $\delta \sim 10 - 100$ curves
in Figure 14 of Schaye et al. (2003), corresponding to overdensities
similar to those of non-linear fluctuations near virialization.  Right
panel: The halo star formation rate for stars of all metallicities (solid)
and $Z = 0$ only with the IGM (dashed) and without the IGM (dotted).
The overall SFR is the same with and without the IGM.  The IGM modestly
inhibits the formation of zero-metallicity stars below $z \simeq 10$,
after the peak caused by disk/halo separation. \label{igm-fig}} 
\end{figure*}

After halo self-enrichment by local star formation, halo-halo
cross-enrichment by metal transport across the IGM is the second possible
mechanism by which halos acquire metals. There are two forms of halo
cross-enrichment - the accretion of metal-enriched IGM gas by an already
virialized halo, and the formation of a halo from gas that has already
been enriched by previous generations of star formation. These avenues
to metal enrichment involve the complicated processes of metal ejection
from virialized halos, their dispersal and mixing into the IGM, and their
incorporation into nearby existing or future halos, for which there is
little observational information at high redshift and no self-consistent
theory.

Previous studies of chemical evolution and IGM metal enrichment vary in
their approach to the problem, with their conceptions of the IGM being the
major discriminant. Oey (2000) treats this problem in the context of an
inhomogeneous chemical enrichment model - metals are transferred across
space to nearby star formation regions, but since these are not termed
``halos'' the ``IGM'' is not explicitly included.  The other approach is
represented by Scannapieco et al. (2003) and Aguirre et al. (2003),
who explicitly consider halo-halo cross-enrichment in the context
of a hierarchical semi-analytic model (the former) or post-processed
hydrodynamical simulations (the latter). If the hierarchical model of
galaxy formation is correct, then the Oey approach, which tracks metal
expulsion, transport, and mixing into nearby star formation regions, is
sensible with only semantic changes to relabel star forming regions into
halos and so on. With this adjustment the two approaches can be considered
on roughly equal footing with respect to observational constraints,
and considered in this light they have advantages and weaknesses. Their
parametric nature allows them to be suitably tuned to match the
available data, but the relatively large number of parameters (10, in
the case of the Aguirre et al. galactic wind model) used to capture the
various physical processes limits their utility to a model like
this one. However, the paucity of observational constraints on the
distributions of metals in the IGM at $z > 5$ means that virtually any
model will have these limitations. It is therefore advisable to make
the model only as complicated as is needed to accomplish its role in
the larger model under development.

This ``minimal model'' tracks the distribution
of gas metallicities in the IGM in terms of a cumulative filling factor
as a function of redshift. For this purpose I use
a very simple parametric model of the IGM, tuned to match the best
available observational constraints at $z = 3$. The model IGM is
calculated independently of the merger trees and so serves as a sort
of ``background'' in which the main model evolves. The model follows
the mass budget of metals expelled into the IGM by virialized halos,
whose space density and mass function are described by the extended
Press-Schechter formalism laid out in Section~\ref{epssection},
using the same star formation law and other related assumptions
used for single halo evolution to calculate an individual halo's
star formation history. Metals produced during this star formation
history are explicitly followed and a mass fraction $f_{esc}^{Z}$ of
these are assumed to escape by one of the viable ejection mechanisms,
most likely supernova-driven winds (Madau, Ferrara, \& Rees 2001) or
halo-halo interactions (Gnedin 1998). Each ``parcel'' of metals driven
from a single halo is then mixed into a time-dependent dilution mass,
given by the same expression as for the ISM (see Eq.~\ref{dil-eq})
but using the IGM parameters listed in Table 1.  The single parameter
set ($f_{esc}^{Z}$, $M_{dil}^{IGM}$, $t_{dil}^{IGM}$) describes the
model uniformly for all halos - the only mass or redshift
dependence to these quantities is imparted indirectly by the star
formation histories.  These are calculated stochastically for evolving
halos in a representative volume of the universe from $z = 70$ to $z
= 0$. By a single redshift, a given halo will have ejected a number
of metal enriched parcels according to its individual star formation
history. Each of these will be mixed into a dilution mass appropriate
to its time since ejection.

At each redshift, the existing metal-enriched regions from all galaxies
is randomly sampled to derive a probability distribution of mass filling
factors for metal-enriched gas as a function of metallicity.  During the
random sampling, all the metal enriched regions from a single halo are
assumed to be perfectly correlated or ``nested''. That is, the smallest
enriched mass associated with a given halo (the most recent one to be
ejected) are assumed to reside entirely within the metal-enriched regions
from previous timesteps. If the random sampling picks one of these less
massive regions, the metallicity for that sample is given by the sum
of all metals from that {\it and previous} metal-enriched parcels with
larger dilution masses. This key assumption has two justifications. First,
it seems reasonable to assume that since galaxies are the sources of
metal enrichment, their successive generations of metal enrichment
are highly correlated, or overlapping, in space and/or mass. Second,
if each successive generation of metal ejection from a single halo is
treated independently, the overall IGM is enriched too quickly to agree
with observations at $z = 3$.

The above procedure is used to calculate a cumulative distribution
of metals, integrated over all halos in a representative volume of
the IGM, in terms of a filling factor $f(Z, z)$.  This is done for
a fixed grid of redshifts from $z = 70$ to $z = 0$ with $\Delta z$ =
0.5. This distribution of filling factors over redshift and metallicity
is used sampled stochastically to find the metallicity of new halos,
or accreted mass, in the halo merger tree chemical evolution model. The
fiducial parameters have been tuned to $f_{esc}^{Z} = 0.05$, $M_{dil}^0 =
5 \times 10^8$ yr, and $t_{dil} = 5 \times 10^9$ yr to match the $\delta
\sim 10 - 100$ filling factor curves at $z = 3$ from the observational
study of intergalactic \ion{C}{4} by Schaye et al. (2003), assuming [C/Fe]
$ = 0.0$. These curves correspond roughly to the non-linear overdensity of
fluctuations near virialization, and so correspond to gas near galaxies
that is likely to be metal-enriched and eligible for later collapse into
new halos.  The effect of the IGM is modest for the highest redshifts,
as seen in Figure~\ref{igm-fig}, which shows the redshift distribution of
metal-free star formation with and without the IGM. The IGM has its
largest effect at $z \lesssim 5$, and its effects are minimized when
disk/halo segregation is included. However, the IGM may become important
later in the model and so it is always used.

After this description of the major ingredients of the chemical evolution
model, it is time to introduce the observational constraints that will
be used to derive specific conclusions about the chemical evolution of
the galaxy and the first generations of stars.

\section{Observational Constraints on Early Chemical Evolution}

The model is constructed to be tested against many different types of
observations, following the theme of relating all observable quantities to
one another. This section describes the various classes of observations,
their power and limitations, and their connection to the model.
Table~\ref{constraint-table} summarizes the basic information and the
constraints they provide on the primordial IMF.

\subsection{Metal-poor Galactic Halo Stars}
\label{empsection} 

These old stars represent the first major observational window into
early stellar populations, as they are perhaps more direct tracers of
the early IMF than are the distant, indirect, and uncertain signatures
of reionization.  Beers \& Christlieb (2005) give a thorough review of
the properties of Galactic metal-poor stars.  I restate here only the most
important features that bear on the present analysis.  These are (1) the
shape of the Galactic halo MDF, (2) the specific Fe-peak element ratios
(especially Zn), (3) the widespread presence of $r$-process elements,
and (4) elevated [C,N,O/Fe] ratios.  This paper adopts the terminology
of Beers \& Christlieb (2005), who label stars according to their
\feh\ like so: \feh\ $<-2.0$ is ``very metal poor'' (VMP), \feh\ $<-3.0$ is
``extremely metal-poor'' (EMP), \feh\ $<-4.0$ is ``ultra-metal poor'' (UMP), \feh
$<-5.0$ is ``hyper-metal poor'' (HMP), and, should any exist, \feh\
$<-6.0$ are ``mega metal-poor'' (MMP).

{\em Galactic halo MDF:} The first key observable derived from the
metal-poor halo stars is the overall metallicity distribution function.
The observed MDF (Ryan \& Norris 1991, hereafter RN91) has three important
properties - a rise from \feh\ $\sim -4$ to a peak at \feh\ $\approx
-1.8$, and a decline at higher metallicity. RN91 state that because
disk stars are difficult to separate conclusively from the halo sample,
stars above \feh\ $= -1.0$ are unreliable. Despite this limitation, the
RN91 MDF is still the most complete in the literature, but will soon be
superseded by larger samples from SDSS and other surveys (T. Beers 2005,
private communication). This study focuses on the \feh\ $< -2$ portion
of the MDF and on constraints that are insensitive to the behavior of
the MDF above the peak at \feh\ $\sim -1.8$.

In addition to the shape of the MDF, the observed limit on the number of
true Population III stars in the Galactic halo is a key constraint on the
IMF of the first stars. The important quantity is \f0, the number of truly
metal-free stars known in the Galactic halo divided by the total number
at [Fe/H] $\leq -2.5$. This quantity is similar to the $F_{III}$ used by
Oey (2002) to examine the implications of Population III non-detections
for Galactic chemical evolution, but with an important difference. Oey
defined $F_{III}$ to be the fraction of truly metal-free stars (1,
to set a limit) divided by the total number in the halo sample, which
was derived from the Beers et al. (1992) survey of 373 halo stars with
\feh\ $< -2.5$ corrected for the Carney et al. (1996) result that 16\%
of all halo stars have \feh\ $<-2.5$. Thus $F_{III} < 4 \times 10^{-4}$.
\f0\ is a slightly different metric that expresses the fraction of
truly metal-free stars relative to the total {\it below \feh\ $<-2.5$}
in the halo sample. This difference in normalization has a two-fold
justification. First, $F_{III}$ may be contaminated by the ambiguous stars
at \feh\ $> -1.0$, some of which may belong in the thick disk. Second,
the details of the model MDF above the peak at \feh\ $\sim -2$ are
sensitive to the method and assumptions by which gas is segregated
from the halo into the disk ($\tau_{disk}$), or by a halo cutoff redshift
($z_{disk}$). Because both observations and theory are uncertain above
the peak of the MDF, it makes sense to define a more robust metric. The
limit on \f0\ is then given by the inverse of the total number of halo
stars with \feh\ $< -2.5$. The 373 stars reported by Beers et al. (1992),
and the 146 Hamburg-ESO survey stars reported by Barklem et al. (2005)
yield \f0\ $\leq 0.0019$.  This limit is used below to provide detailed
constraints on the IMF at $Z < Z_{crit}$.

{\em Iron-peak elements (Cr -- Zn):} These elements are produced
in explosive events and so trace nucleosynthesis by massive stars.
Large samples presented by McWilliam et al. (1995) and Carretta
et al. (2002) found that the Fe-peak ratios to Fe, i.e. [Zn/Fe],
qualitatively change their behavior at \feh\ $ \simeq -3$, but the
more recent high resolution studies by Cayrel et al. (2004) and Cohen
et al. (2004) have found lower scatter and a smooth rather than abrupt
change in the abundance ratios at \feh\ $\simeq -2$ to $-4$. In any
case, the overall abundance pattern is consistent across the different
studies. Theoretical yields of iron peak elements from metal-free stars
have been calculated by a number of groups (e.g., HW02 for VMSs; Umeda \&
Nomoto 2005 for core-collapse SNe). These yields are sensitive to the
energy, rotation, mass cut, and asymmetry of the supernova and to the
pre-SN stellar properties. The models can be tuned to match individual
observed abundance patterns by varying these assumptions. Because of these
sensitivities and the remaining disagreement between different codes and
investigators, and the intrinsic difficulty of calculating yields {\em
a priori}, this study does not rely on detailed comparisons between
theoretical yields of core-collapse SNe and observations of metal-poor
stars to constrain the IMF, except in one case which makes use of a unique
and widely agreed circumstance for zinc (see \S~5.3).  Instead this study
has the ultimate goal of extracting intrinsic yields directly from the
data, guided by models of chemical evolution.

{\em $r$-Process elements (A $>$ 60):} These elements are produced
by rapid neutron captures in hot, dense, neutron-rich explosive
events. Although the exact physical sites are still uncertain, the
proposed mechanisms are all associated with massive stars in the range
$M = 8 - 40$~\msun\  and may be isolated to $8 - 12$ \msun\ (Truran et
al. 2002).  The existing samples of metal-poor stars show $r$-process
elements down to \feh\ $\sim -3.5$ (McWilliam et al. 1995; Burris et al.
2000; Barklem et al. 2005). The mean [$r$/Fe] is similar to the solar
value at all \feh, but with scatter steadily increasing to 2 dex at \feh\
$\sim -3$. The relative abundances (i.e., [Eu/Ba]) are similar enough
to the solar ratios to suggest that the details of the r-process
are unchanged over more than 3 decades of metallicity (Truran et
al. 2002). TVS04 argued that the presence of r-process elements in the
EMPs effectively rules out the strong VMS hypothesis because PISNe do
not experience the conditions in which these elements are created. For
this reason the strong VMS hypothesis is not treated in detail here.

{\em Primary elements (C, N, O):} Some UMP stars are highly enriched
in carbon relative to iron. This group includes the two most iron-poor
stars known, HE 0107-4342 (Christlieb et al. 2003) and HE 1327-2326
(Frebel et al. 2005), both of which have [C/Fe] $ > 2.0$.  These stars
appear to be {\em iron-poor}, rather than {\em metal-poor}, and
therefore less chemically primitive than \feh\ indicates.  Carbon is
likely the dominant coolant that determines the IMF in the early ISM
(Bromm \& Loeb 2003) and the nuclear burning catalyst that determines
by its absence the unusual evolution and radiation of the first stars
(TSV03). These factors favor C instead of Fe as the ``reference element''
for finding chemically primitive stars in the local universe. No attempt
is made here to explain these unusual stars on their own, and they are
not included in the MDF sample.

The features of the metal-poor stars discussed here provide many
interesting constraints on the history of chemical evolution in the Galaxy
and on the first stars. Some of these constraints, such as those provided
by \f0\ and the large r-process enhancements down to \feh\ $\sim -3$,
are quite robust and are not sensitive to the model. By contrast, making
detailed use of the Fe-peak ratios requires some detailed knowledge of
intrinsic SNe yields. These constraints are less robust because they are
sensitive to the details of the specific theoretical model from which
they are derived. This model is directed at extracting SNe yields directly
from the data, so it is best to avoid constructing a model which depends
on pure theoretical yields for its connection to the data. Nevertheless
there are some reasonable constraints that can be drawn from the specific
metal-abundance patterns and theoretical yields, where these are thought
to be sound. These results are presented below with appropriate caveats.

\subsection{Reionization}
\label{reionsection}

The reionization of the IGM at high redshift is a key indicator of cosmic
star formation activity at $z \gtrsim 6$. There are two major data that
guide models of reionization. First, the Gunn-Peterson effect has been
detected in the spectra of $z \gtrsim 6$ quasars, indicating that the
epoch of reionization ended near $z = 6$. Second, the unexpectedly high
electron-scattering optical depth to the cosmic microwave background
found by {\it WMAP}, $\tau_{es} = 0.17^{+0.08}_{-0.07}$, suggests
that reionization began quite early and may have had a complicated
history.  Most of the ionizing photon budget that reionizes the IGM
is thought to be produced by massive stars, as the number density of
bright QSOs at that time is probably not sufficient (Loeb \& Barkana
2001). However, there is not widespread agreement about the details
of when reionization began, how it proceeded, and how galaxies of
different mass and metallicity contributed to the process. A number of
recent studies have approached this problem with semi-analytic models of
the growth of structure and the evolution of cosmological H~II regions
into the IGM with varying assumptions and different results (Haiman \&
Holder 2003; Cen 2003; Venkatesan, Tumlinson, \& Shull 2003). All these
models have two common conclusions: (1) that the overall efficiency of
converting baryonic mass to ionizing photons must be higher than for a
metal-free stellar population with a Salpeter IMF between 1 - 100 \msun,
and (2) that ``minihalos'' with $T_{vir} \leq 10^4$ K, where most of
the high-redshift baryons reside, must participate in reionization if
the {\it WMAP} $\tau_{es}$ is to be reproduced.

TVS04 studied the ionizing properties of different IMFs and argued that
a standard IMF deficient in low- and intermediate-mass stars could be
just as efficient as VMSs at reionizing the IGM.  In a stellar population
synthesis, the quantity most relevant to reionization is the total number
of ionizing photons produced per baryon locked into stars. If integrated
over the lifetimes of the stars, this number is called $\gamma_0$.
According to Figure 2 of TVS04, this quantity peaks at $\simeq 120$
\msun\ for metal-free stars, and then declines at higher mass.  For a
Salpeter IMF from 0.5 - 140 \msun, $\gamma_0 = 17000$.  This value is
generally not sufficient to reproduce the $\tau_{es} \geq 0.10$, while
TVS04 showed that a similar IMF truncated at $M = 10$ \msun\ yields
$\gamma_0 = 65000$ and can produce $\tau_{es} = 0.11 - 0.13$ without
extreme assumptions for star formation efficiency ($f_{*}$) or the
Lyman continuum escape fraction ($f_{esc}$).  The TVS04 underlying
structure formation model is based on the same extended Press-Schechter
formalism, and the overall assumptions about star formation efficiency
are comparable.  Thus for this study I simply interpolate between the
different IMF cases presented in TVS04 and conservatively require that
$\tau_{es}$ achieve the minimum value consistent with the {\it WMAP} data
at $1 \sigma$ confidence, $\tau_{es} \geq 0.10$.  In the TVS04 models,
this requires $\gamma_0 \geq 34000$.  These results appear in \S~5.
A full treatment of the detailed interrelationships between chemical
evolution and reionization is enabled by this framework and certainly
warranted, but it is too involved to take place here.

\section{Constraints on the Primordial IMF}

The IMF is, perhaps, the most fundamental feature of a stellar population,
and therefore it is the most important fact of chemical evolution.
For the same reasons, the IMF has many observable consequences.
Some of these are used here to constrain the IMF of the first stars.
This discussion starts with the most robust observational results and
the strongest connections between data and theory, and then proceeds
to the more model-dependent connections and looser observational
constraints.  The basic constraints and their results are summarized
in Table~\ref{constraint-table}.

\subsection{Constraints on the IMF from the Galactic Halo MDF and
Population III Stars}

\begin{figure}
\centerline{\epsfxsize=\hsize{\epsfbox{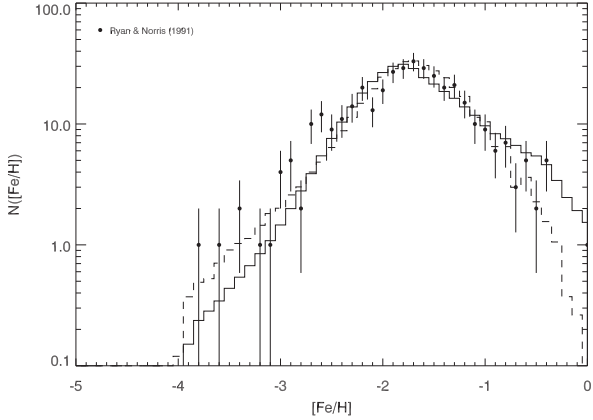}}} 
\figcaption{Comparison of theoretical MDFs for the fiducial
\tree{5}{11}{3}{8} tree (solid) and the fiducial \tree{5}{11}{4}{8} tree
(dashed) to the RN91 MDF for a log-normal IMF (10, 1.0) below $Z_{crit}
= -4.0$.  Both the model and observed MDFs are uncertain above \feh\
$\sim -1.0$. \label{mdffig}} \end{figure}

The Galactic halo MDF is the first key result of the model.  Good fits
to the peak of the MDF are obtained for $\tau_{disk} = 8 \times 10^7 -
3 \times 10^8$ yr for $T_{vir} = 10^3$ and $10^4$ K, respectively,
when no halo cutoff redshift $z_{disk}$ is assumed.  The MDF can
be fitted with reasonable parameters with or without minihalos (see
Section~\ref{minihalosection}), leaving the minimum virial temperature
uncertain.  These fits work well below \feh\ $<-1.0$, above which the MDF
is declared unreliable by RN91.  The fit in this region can be improved
by the introduction of $z_{disk} = 8 - 12$ (see \S~\ref{diskhalosection})
with no effect on \f0.  The best-fit MDFs for \tree{5}{11}{3}{4} and
\tree{5}{11}{4}{8} trees are shown in Figure~\ref{mdffig}.  This basic
agreement between the observed and model MDFs provided encouraging
evidence that the overall the model is an accurate description of early
Galactic chemical evolution.

Another important quantity derived from the MDF is \f0, the number
fraction of true ``Population III'' stars seen in the Galactic halo.
The first key result from \f0\ is that the qualitative change in the IMF
expressed by $Z_{crit}$ {\em must} exist. If no $Z_{crit}$ is employed,
the IMF is the same at all metallicities and \f0\ $\gtrsim 0.1$ for IMFs
ranging over (1.0, 1.0) to (100, 2.5).  For a Miller-Scalo IMF (0.1, 1.57)
at all metallicities, the model gives \f0\ = 0.48 for standard ISM mixing
parameters and 0.25 in the instantaneous mixing limit. A Salpeter IMF with
minimum mass 0.5 \msun\ gives \f0\ = 0.45 for standard mixing and 0.23
for instantaneous mixing. These values are clearly excluded by the data,
and so there must be an evolution in the IMF that inhibits the formation
of low-mass stars below some finite metallicity.  The observations {\it
require} a $Z_{cr}$, although its specific value cannot be constrained
below the most metal-poor star in the sample (see below).

\begin{figure}
\centerline{\epsfxsize=\hsize{\epsfbox{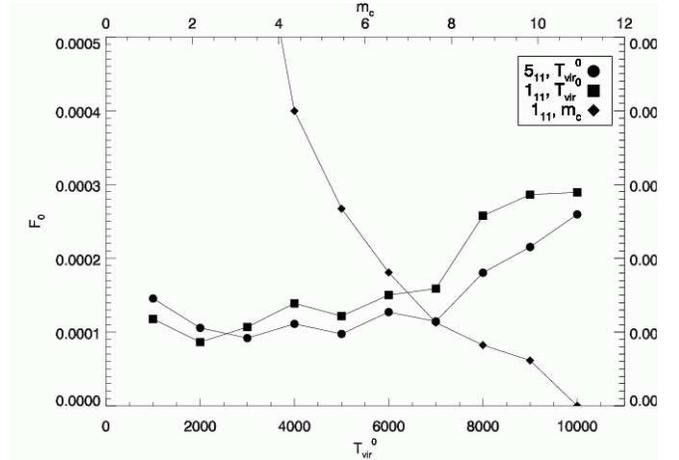}}} 
\figcaption{Variation of \f0\ with $T_{vir}^{0}$ for trees of $M_h = 1$
and $5 \times 10^{11}$ \msun\ (filled circles and squares, respectively)
and an IMF with $m_c = 10$, $\sigma = 1.0$. The filled diamonds show
\f0, divided by 10, for variations in $m_c$ as shown on the top axis,
and with fixed $\sigma = 1.0$. Variations due to statistical noise and
systematic uncertainties in $M_h$ and $T_{vir}^0$ are overwhelmed by
variations with IMF parameters.  \f0\ is therefore a robust indicator
of the IMF. \label{covar-fig}} \end{figure}

For the fiducial $Z_{crit} = -4.0$, \f0\ varies dramatically across
the range of IMF parameters and so provides a critical constraint on the
IMF at the low-mass end.  Figure~\ref{covar-fig} shows the variations
of \f0\ with total halo mass, $M_h$, and minimum virial temperature
$T_{vir}^{min}$ (see Section~\ref{minihalosection} regarding the
importance of minihalos with $T_{vir} \leq 10^4$ K), with the curve
for \f0\ plotted on the same scale but divided by ten. Variations in
\f0\ with IMF overwhelm uncertainty in the other parameters and make
\f0\ a robust constraint on the IMF that should only improve with new
samples of metal-poor stars (until a truly metal-free star is directly
detected).  Of all the constraints used here, this one is least subject
to astrophysical uncertainties and effectively has no model dependence.
These results are displayed in panel A of Figure~\ref{imffig} over
a large range of IMF parameters and the fiducial \tree{5}{11}{3}{8}
tree. The line marking \f0\ $= 0.0019$ passes through (4.0, 0.6) 
to (8, 1.5) before passing out the top of the frame.  Thus \f0\ provides
a strong requirement that the initial IMF is top-heavy, with $m_c
\gtrsim 6$ \msun\ and mean mass of $\langle M \rangle \gtrsim 8$ \msun. 
In this parameter space, the Miller-Scalo Galactic IMF has $m_c = 0.1$,
$\sigma = 1.57$, and $\langle M \rangle \approx 0.64$ \msun. Thus our
strongest constraint on the IMF, \f0, implies an IMF with mean mass at
least 10 times that found in the present day.

The MDF may also contain evidence of the expected ``critical metallicity''
at which interstellar cooling becomes dominated by metal lines. For
a single $Z_{crit}$, this signature would be expected to take the
form of a noticeable break in the MDF at a single \feh. Inspection
of Figure~\ref{mdffig} shows that such a break is not apparent in the
RN91 MDF, which has only 10 stars at \feh\ $\leq -3$.  However, there
could still be an indication of a statistically significant deficit of
low-\feh\ stars relative to expectations from the model. This
is done by constructing the {\it cumulative} metallicity distribution,
integrating up from zero metallicity to \feh\ $= -2$. For this test,
the model MDF is normalized to have the same total number of stars at
\feh\ $ \leq -2$ as the RN91 MDF (125 stars).

The results of this test appear in Figure~\ref{zcrit-fig}, which shows
cumulative MDFs for RN91 and the fiducial \tree{5}{11}{3}{8} tree and
four different values of $Z_{crit}$. A model with $Z_{crit} = 10^{-5.0}$
overproduces EMP stars, while $Z_{crit} = 10^{-3.5}$ underproduces them.
For $Z_{crit} \leq 10^{-4.5}$, the model MDF converges to a single curve
because adding stars with extremely low \feh\ adds very little to the
cumulative total.  The MDF is matched reasonably well by $Z_{crit} \simeq
-4.0$, or roughly the lowest metallicity present in the RN91 MDF.  The MDF
appears smooth down to \feh $= -3.8$, and this smooth distribution can
be explained by the same model parameters and IMF that match the other
observables. Thus it seems that the observations require $Z_{crit} \leq
10^{-3.8}$.  Indeed, the IMF and $Z_{crit}$ are intimately connected,
because the number of low-mass stars below $Z_{crit}$, that can
contribute to the cumulative distribution, depends on the IMF at $Z <
Z_{crit}$. Unless a clear-cut break is seen in future updates to the
observed MDF, $Z_{crit}$ will probably not be constrained in this fashion.
However, it can be constrained along with all the other model properties,
i.e., if an IMF can be independently determined by \f0\ and reionization,
$Z_{crit}$ can be inferred.  This test would also require much better
statistics in the low-\feh\ bins, where currently there are only a
handful of stars.  Because of the dearth of data below \feh\ $\lesssim
-3$, it seems the critical metallicity for normal star formation has
not yet been detected.

The rigid limit at a single \feh\ is a direct consequence of the assumption
that there is a single $Z_{crit}$.  Models by Santoro \& Shull (2005)
suggest that the critical metallicity can depend on both the particular
mix of metals (Si/C rich vs.~Fe rich) and the initial temperature and
density conditions of collapse. These additional factors can blur out the
single $Z_{crit}$ assumed here, and may in fact yield a range of critical
metallicities that extends above the lowest-\feh\ star seen in the halo.
Future work will endeavor to incorporate a broader range of assumptions
about $Z_{crit}$ and connect them to observations and detailed theory.

\subsection{IMF Constraints from Reionization} 

\begin{figure*}
\centerline{\epsfxsize=\hsize{\epsfbox{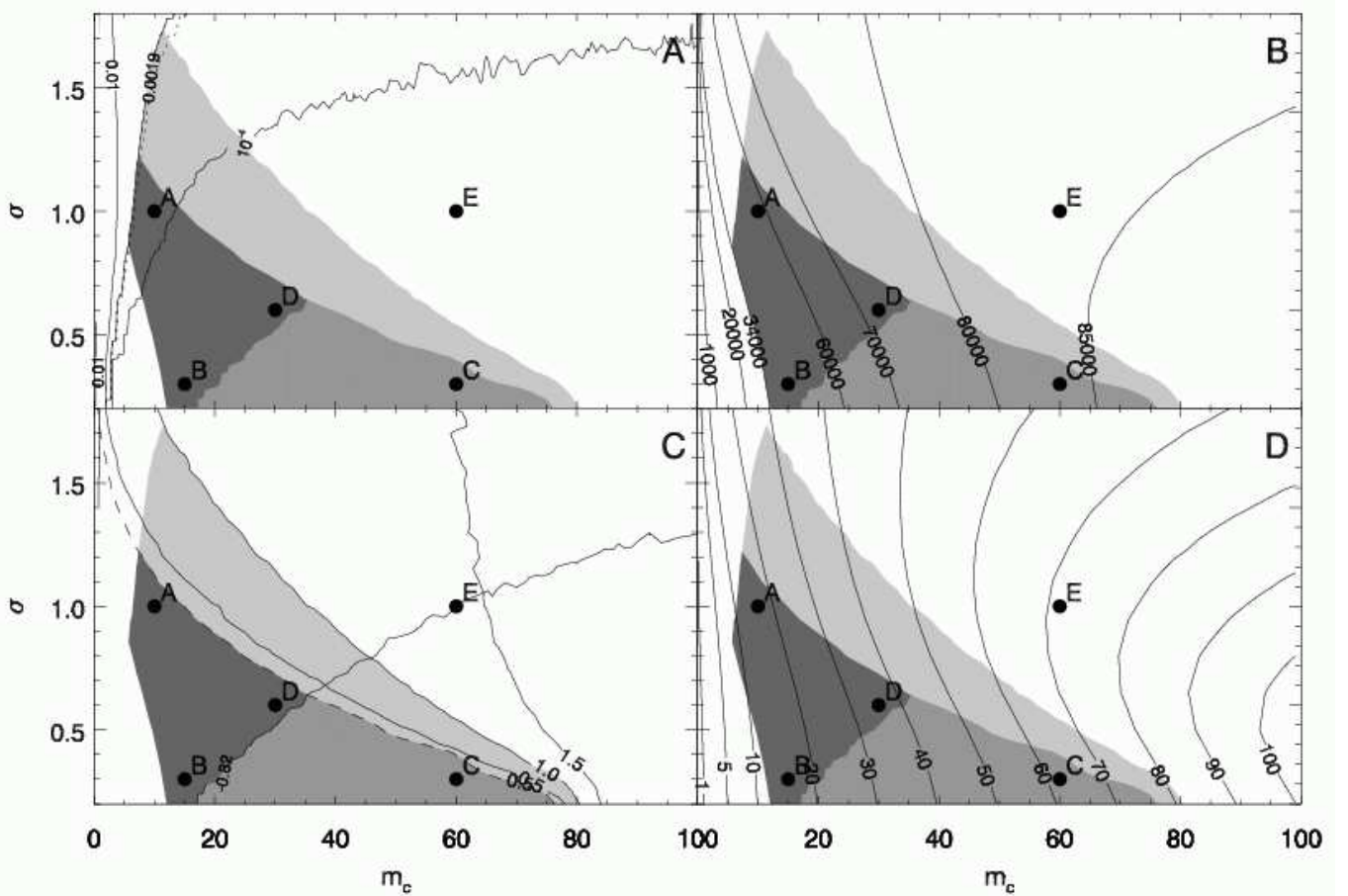}}} 
\figcaption{Observational constraints on the parameters for a primordial
IMF. Each panel shows contours for an individual constraint, while the
shaded regions that meet all constraints are the same in all panels.
Panel A: Contours of \f0\ for log-normal IMF parameters. \f0\ $\leq
0.0019$ requires $m_c \gtrsim 4$ \msun.  Panel B: Contours of constant
ionizing efficiency $\gamma _0$ for $Z = 0$ stellar atmosphere models
from TSV03.  Panel C: Solid contours show constant [Zn/Fe] from 8 -
40 \msun\ that is required for the given IMF to yield average [Zn/Fe]
= 0.3 in the metal-poor stars from the Cayrel et al. (2004) sample.
Dashed contours show constant iron production ratio from VMSs, $F_{VMS}$.
Above the contours [Zn/Fe] = 1.0 and $F_{VMS} = 0.9$, VMSs dominate the
Fe production so much that unsuitably large Zn excess is required from
massive stars. Also shown with a dashed line is the 
contour of constant $F_{VMS} = 0.5$, or one-half of the total iron budget
coming from VMSs. See text for discussion.  Panel D: Mean stellar mass of
a log-normal IMF with the given parameters. This quantity is a derived
output of the model and not an independent constraint.  \label{imffig}}
\end{figure*}

The second major constraint on the IMF is that the ionizing photon budget
must be sufficient to provide a high optical depth to reionization,
$\tau_{es} \geq 0.1$. To meet this constraint, the TVS04 models suggest
that $\gamma_0 \geq 34000$ is needed. This is a somewhat model-dependent
constraint on the IMF, because it involves a number of poorly understood
astrophysical parameters, such as the escape fraction of ionizing
radiation from early galaxies, and the disposition of ``minihalos'' with
$T_{vir} < 10^4$ K. As argued in \S~\ref{reionsection}, $\gamma_0 \geq
34000$ is needed to give $\tau_{es} \geq 0.1$, the smallest value within
the $1 \sigma$ confidence interval of the first-year {\em WMAP} data.
Panel B of Figure~\ref{imffig} shows contours of $\gamma_0$ in the
IMF parameter space, using the mass-dependent $\gamma_0$ results in
Figure 2 of TVS04.  Clearly $\gamma_0 \geq 34000$ requires an IMF skewed
to high mass, such that the same general range of low $m_c$ and small
$\sigma$ excluded by \f0\ is also excluded by $\gamma_0$. Together,
\f0\ and $\gamma_0$ constrain the IMF to $m_c \geq$ 6 \msun. IMFs
with much higher $\gamma_0$ are permitted, including very efficiently
ionizing IMFs that could give $\tau_{es} \gtrsim 0.14$. In this sense
the requirement that $\tau_{es} \geq 0.10$ (1$\sigma$ confidence from
{\it WMAP}) is conservative. Meeting the $\tau_{es}$ constraint seems
to require at least some star formation in minihalos with $T_{vir}
\leq 10^4$ K, which complicates some of the other IMF limits (see also
Section~\ref{minihalosection} for more discussion).

\subsection{IMF Constraints from Detailed Nucleosynthesis} 

Additional, and more model-dependent, constraints on the IMF come from
the detailed abundances seen in metal-poor stars. As argued in TVS04,
the presence of the r-process elements in the EMPs do not allow the
early budget of iron to be provided exclusively by PISNe, which have
no r-process (HW02; TVS04).  TVS04 also argued that the early budget of
Fe could not be dominated by the peculiar nucleosynthetic signature of
VMSs, which show a strong odd-even effect and deficits in some Fe-peak
elements (particularly Zn). That paper did not, however, attempt to place
quantitative constraints on the Fe budget that could have come from VMSs.
This boundary can be defined precisely only if detailed models using
Type II and PISNe yields are fitted with the observed samples for
the full range of IMF parameters, using some type of goodness-of-fit
metric. However, this analysis would require a complete ``basis set''
of mass- and metallicity-dependent yields for metal-free Type II
SN, which does not exist (especially for core-collapse SNe models,
which do not spontaneously explode and have many tunable parameters).
Because extracting detailed empirical yields directly from the data
on metal-poor stars is a goal of the final model, it is best to avoid
direct reliance on the theoretical yields where possible.  Despite this
desire, there is probably no reasonable alternative for constraining the
upper right corner of the IMF parameter space, where VMSs overwhelmingly
dominate the mass and iron budget. This portion of parameter space easily
meets the \f0\ and $\gamma_0$ limits, leaving nucleosynthesis and future
high-redshift galaxy searches as the remaining possible constraints.
For the purposes of this study it is sufficient to establish some
conservative upper limit to the contributions of VMSs to early chemical
evolution. This constraint implicitly assumes that all SNe yields are
retained in their parent halos. This assumption is likely to be violated
in minihalos with $T_{vir} \lesssim 10^4$ K, so if they prove necessary
to match the overall dataset, these limits will need to be reevaluated.

When PISNe yields are compared against the stellar abundance data,
either by integration over an IMF (HW02) or for individual masses
(TVS04), the most discrepant element is zinc. PISNe produce very little
Zn in an $\alpha$-rich freeze-out, and that only in the upper mass
range. Thus high Zn yields must be produced from $8 - 40$ \msun\ for
the IMF-weighted mass yields to match the data, which for the Cayrel et
al. (2004) sample shows $\langle [{\rm Zn/Fe}] \rangle = 0.3$. A possible 
constraint comes from asking what typical Zn yield is needed for 8 -
40 \msun\ to compensate for PISNe and match the data with a given IMF.
The solid contours in panel C of Figure~\ref{imffig} mark the [Zn/Fe]
excess that is needed, uniformly across the range 8 - 40 \msun, to
overcome the strong [Zn/Fe] deficit of PISNe and still give a mean
[Zn/Fe] $\approx 0.3$ for stars with \feh\ $\leq -2$.  IMFs in the
upper right corner require very high Zn excess, [Zn/Fe] $\gtrsim 1.0$.
To constrain the IMF, I adopt [Zn/Fe] $\leq 1.0$ as a conservative
limit.  This constraint on the Fe budget from VMSs places a constraint
on how top-heavy the IMF can be that is complementary to \f0.  The IMF
cannot occupy any region above the line which roughly runs from (20,
1.5) through (35, 1.0) to (60, 0.5).  This still leaves a lot of
parameter space, but has at least provided a key constraint at the upper
end, complementary to the \f0\ constraint at the lower end of the IMF.

This [Zn/Fe] limit is called conservative for two reasons. First, this
Zn excess is higher than that produced in typical published models of
core-collapse supernova nucleosynthesis.  Umeda \& Nomoto (2002) show
that with no mixing and/or fallback to enhance the elements above Fe,
[Zn/Fe] $\lesssim 0.0$. Their models that include mixing and fallback
with the explicit purpose of better matching the Fe-peak elements (Cayrel
et al. 2004) have [Zn/Fe] $ = 0.3 - 0.5$, with a maximum of $+0.5$
produced by a supernova that explodes with $E = 30 \times 10^{51}$ erg
(Umeda \& Nomoto 2005). The zero-metallicity models of Woosley \&
Weaver (1995) all have [Zn/Fe] $\lesssim 0.0$. Thus even assuming
[Zn/Fe] $=0.5$ for all 8 - 40 \msun\ SNe strains the existing models
to their limits.  In this sense, placing the limit at [Zn/Fe] $\leq
1.0$ is conservative.  Second, [Zn/Fe] $\leq 1.0$ for 8 - 40 \msun\
limits the Fe budget contributed by PISNe rather generously to $f_{VMS}
\leq 90$\%.  This limit is called generous because it is probably more
than can reasonably be accommodated by the observed abundances in the
metal-poor stars.  HW02 note that a power-law IMF with Salpeter slope
(which produces $\sim 50$\% of its Fe with VMSs given the adopted SN mass
ranges) has too pronounced an odd-even effect. This tighter constraint
is displayed in a dashed contour in panel C of Figure~\ref{imffig}
and adopted as a more realistic constraint on the IMF.  A 10\% percent
contribution from VMS is probably still more realistic given their
pronounced odd-even effect.  There are two reasons for setting this limit
conservatively. First, deriving more quantitative, and more restrictive,
constraints on $F_{VMS}$ based on some goodness-of-fit-test would
require undesirable reliance on the detailed theoretical yields.
Second, as noted by HW02, Zn may have an alternative production site
beyond the $\alpha$-rich freeze-out at the higher masses. There may
be a neutrino-driven wind that could leave an important contribution
to Zn production and permit lower [Zn/Fe] from massive stars (8 -
40 \msun). This feature has not been modeled extensively for massive
stars or VMSs, so it remains speculative. Of course, this would leave
the severe odd-even effect, but this is less quantifiable. In light of
these two motivations the conservative limit on [Zn/Fe] seems reasonable.
Thus weak-VMS IMFs are permitted by the existing data and conservative
assumptions about intrinsic yields.

There is an additional constraint from detailed nucleosynthesis that is
more speculative but potentially even more powerful than the observed
ratios of iron-peak elements for constraining the IMF - the frequency
of r-process enrichment observed in metal-poor stars. This constraint is more
speculative because it relies on an assumption that r-process elements
are uniquely associated with a narrow range of the IMF, between 8 and 12
\msun. It is potentially more powerful for exactly that same reason.
The basic idea is possible because the r-process elements (here
represented by barium) cannot be produced in PISNe, and are thought to
be associated with core-collapse SNe in the range 8 - 12 \msun. Thus we
can use the observed frequency of r-process enrichment at \feh\ $< -2$
as leverage on the shape of the IMF.

\begin{figure}
\centerline{\epsfxsize=\hsize{\epsfbox{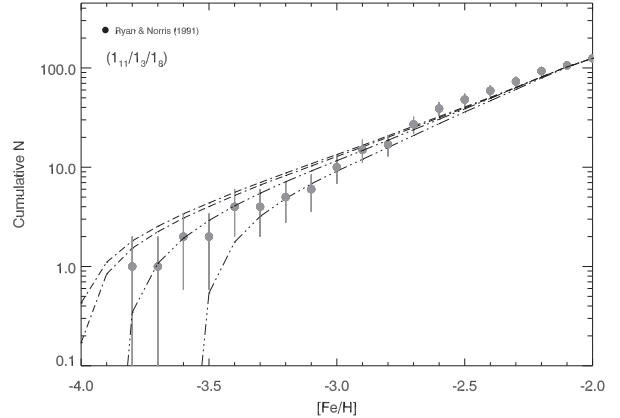}}} 
\figcaption{Cumulative MDF for RN91 and a \tree{1}{11}{3}{8}
tree and $Z_{crit} = -5.0, -4.0, -3.8, -3.5$ marked with lines, from top.
$Z_{crit}$ appears to be $< -3.5$ and fits best for $Z_{crit} = -3.8$
(the most metal-poor star from RN91 has \feh\ = -3.7) but is otherwise not
well constrained by the few points below \feh\ $= -3$.\label{zcrit-fig}}
\end{figure}

This simple model for r-process enrichment that follows in the spirit
of the phenomenological model of Fields, Truran, \& Cowan (2002).
The sample starts with the HERES sample of Barklem et al. (2005), and
is first culled to include only their 228 stars from the HE survey.
I include only stars with with [Fe/H] $<-2.5$, below which it is
believed that all Ba comes from the r-process (Truran et al. 2002).
In this remaining sample of 146 stars, 119 (82\%) show detected Ba at
[Ba/Fe] $ > -2$. To test this against model IMFs, chemical evolution
histories for low-metallicity stars are constructed, allowing for 10 SN
progenitors (see \S~6.1). Iron yields are assumed as given above for the
Type II (8 - 40 \msun) and PISNe (140 - 260 \msun) mass ranges.  Over the
range 8 - 12 \msun, SNe are assumed to produce Ba in the highest observed
pure r-process ratio relative to Fe, [Ba/Fe] = 1.5. This high ratio,
which is only observed for rare (r-II; Barklem et al. 2005) objects,
represents an extreme high value for the r-process Ba yield of a single
SNe and so helps set a conservative limit on the iron-rich/r-process
poor contribution at the high end of the IMF.

The results of this test appear in Panel C of Figure~\ref{imffig}.
The contour labeled ``0.82'' separates IMFs which have a suitable ratio
of r-process to iron progenitors (above the contour) from those that
produce too much iron and not enough r-process enrichment. The parameter
space excluded by this test is reproduced in the medium gray shade in
the other panels.

This is the most model-dependent of the constraints placed on the IMF,
because it assumes that r-process enrichment is produced only by stars
with 8 - 12 \msun.  This mass range is favored by theoretical models of
the r process in supernovae (Woosley et al. 1994; Truran et al. 2002) but
is still unconfirmed by observation. If the r process occurs in SNe across
the full range (8 - 40 \msun), r-process progenitors are quite numerous
and push the 82\% constraint out of the interesting IMF parameter space.
For this reason this test is only a tentative constraint and continue
with an IMF test case that lies in the region excluded by this uncertain,
but potentially very powerful test.

These constraints from detailed nucleosynthesis exploit the unusual
behavior of PISNe from very massive stars with $140 - 260$ \msun. At
the high mass end, these stars eject up to 40 \msun\ of iron and
correspondingly large masses of other detectable elements (HW02). However,
as discussed below in \S~5.5, if these stars form predominantly in
minihalos with $T_{vir} \lesssim 10^4$ K, they may also disrupt their
parent halos with energetic supernovae, and therefore not leave their
nucleosynthetic products eligible to enrich later generations of star
formation. While minihalos are not required to match the observed
MDF and \f0\ constraints, they may in fact be required to form stars
efficiently to meet the reionization constraints of {\em WMAP} or future
CMB results. This section has implicitly assumed that these VMSs form in
$T_{vir} > 10^4$ K halos in addition to minihalos, and that their ejected
metals remain bound. Thus the nucleosynthesis constraints on the upper
end of the IMF, discussed here, and the reionization constraints may not
be achievable simultaneously.  Minihalos are discussed more thoroughly
below in Section~\ref{minihalosection}.

\begin{figure}
\centerline{\epsfxsize=\hsize{\epsfbox{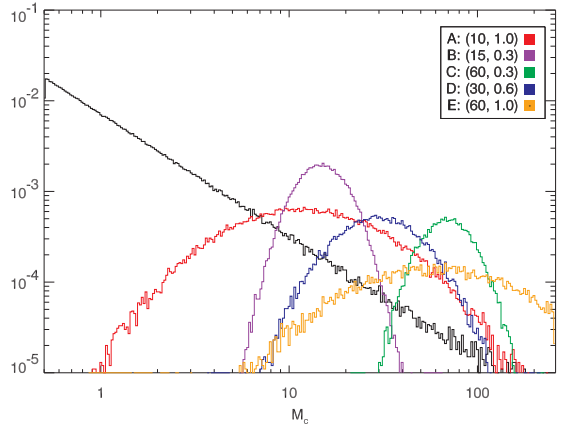}}} 
%\plottwo{../figures/real_imf.eps}{../figures/pop3-sfr.eps}
\figcaption{The possible log-normal IMFs drawn from Figure~\ref{imffig},
compared with a power-law IMF of Salpeter slope, $\alpha = -2.35$, all
normalized to have the same total mass $10^6$ \msun. The best-fitting
IMFs are all sharply peaked in the range 12 - 70 \msun. See the discussion
in the text for details. \label{realimffig}} \end{figure}

\subsection{IMF Test Cases} 

The various constraints in combination suggest an IMF for the first
stars that is restricted to the dark shaded area in Figure~\ref{imffig}.
This IMF is restricted to $m_c = 6 - 35$ \msun, with $\sigma \approx 0.3
- 1.2$ and mean mass $\langle M \rangle = 8 - 42$ \msun.  Within this
permitted range, there are not yet detailed constraints that can
discriminate the parameters still further. I define five IMF test
cases for further examination of their detailed properties and
consequences. These are labeled IMF A (10, 1.0), B (15, 0.3), C (60,
0.3), D (30, 0.5) and E (60, 1.0). These represent, respectively, two
low-$m_c$ models which are only as top-heavy as they need to be, but
with different widths, a high-mass model sharply peaked at $\langle M
\rangle = 70$ \msun\ (perhaps excluded by r-process abundances), a model
intermediate between A, B, and C, and then E, an extreme top-heavy IMF
useful for exploring the the importance of feedback on minihalos and
the feasibility of the [Zn/Fe] constraint. These test cases are shown in
the left panel of Figure~\ref{realimffig} compared against the power-law
Salpeter IMF.

\begin{figure*}
%\centerline{\epsfxsize=0.7\hsize{\epsfbox{f9a.eps}}} 
%\centerline{\epsfxsize=0.7\hsize{\epsfbox{f9b.eps}}} 
\plottwo{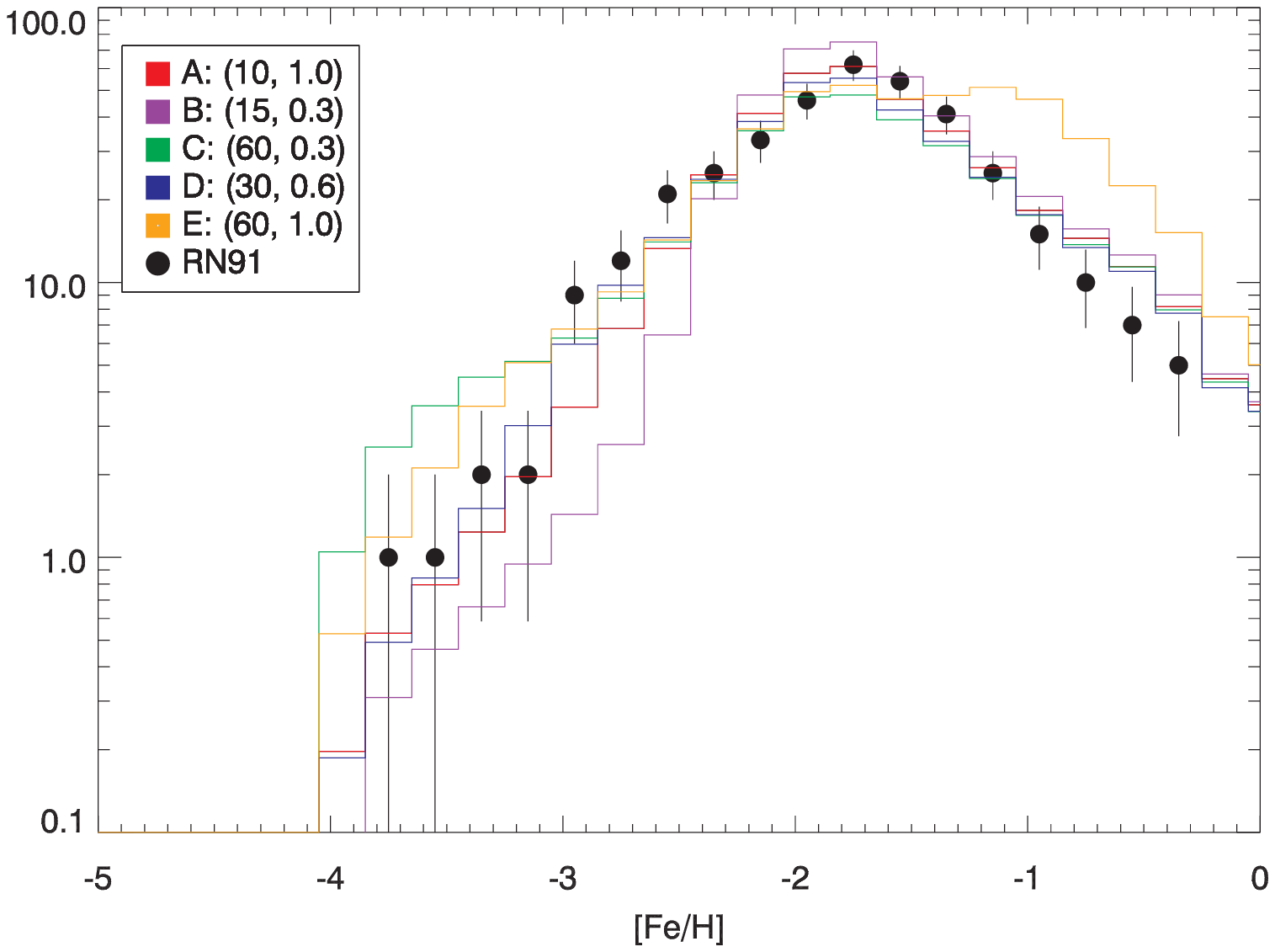}{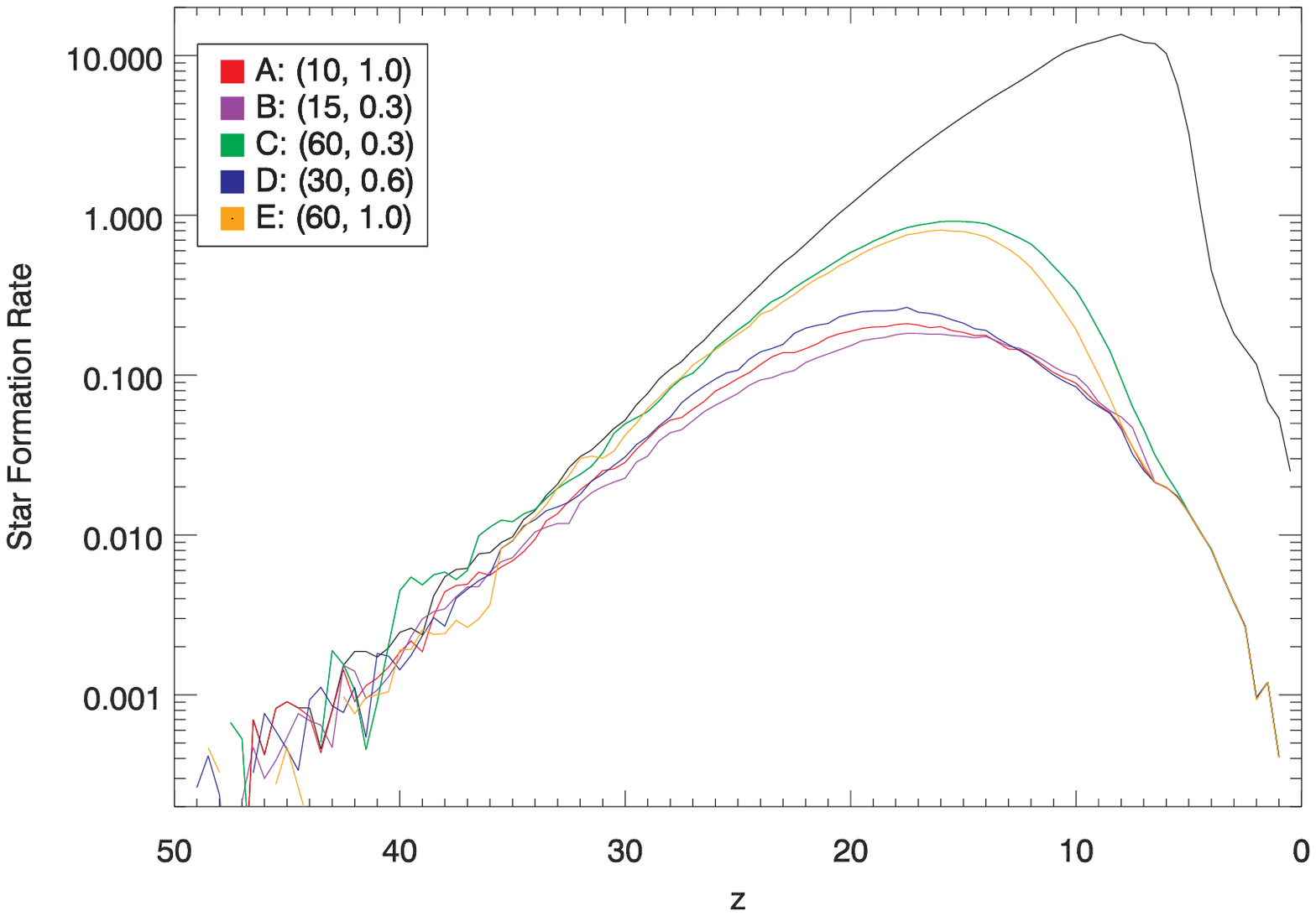} 
\figcaption{Left panel: Comparison of MDFs for the five IMF test cases,
normalized to the same total number of stars as in the complete RN91
MDF. Cases A and D yield reasonable fits, but cases C and E 
show a strong ``bump'' feature just above $Z_{crit}$ that is not seen
in the data, while case B is deficient in EMP stars.  Models sharply
peaked with $m_c = 10 - 30$ \msun\ are favored by the overall dataset.
Right panel: Comparison of metal-free SFRs for these different IMFs,
which may be distinguishable in the future from the level of metal-free
star formation activity observed at high redshift.  Model IMFs C and E
have the highest mass in the 50 - 140 \msun range, so they results in
the highest overall $Z = 0$ SFR. The black curve shows the overall SFR
(at all metallicities). This quantity is not sensitive to the IMF and
is represented here by IMF case A.  \label{imfcasefig}} \end{figure*}

The MDF provides a possible test of the IMF cases. The results of these
comparisons appear in Figure~\ref{imfcasefig}. IMFs A and B provide
the overall best fit. IMF B has a strong deficiency of VMP and EMP
stars relative to the RN91 MDF, while IMFs C and E have an excess. This
``bump'' feature is caused by the strong deficiency of stars that produce
a nucleosynthetic signature (8 - 40 \msun) relative to those that do
not (40 - 140 \msun).  Both of these IMFs (C and E) are potentially
excluded with detailed nucleosynthetic results presented above and
they also produce poor fits to the overall MDF.  For these reasons we
disfavor these IMFs.  However, these MDF results are not decisive for
determining the IMF because of the poor statistics in the data below \feh\
$\sim -2.5$.  Improvements in the sample size expected from future large
surveys of Galactic halo metal-poor stars have the potentially to tighten
IMF constraints based on the Galactic halo MDF.

Different star formation histories for metal-free stars can also 
help discriminate the IMF test cases.  An IMF that is more top-heavy,
and which places more mass into the $40 - 140$ \msun\ range of stellar
mass that does not produce a nucleosynthetic signature, will take longer
to produce a fixed overall enrichment and so should form metal-free
stars for a longer time in each halo. This effect could potentially be
detected at high redshift. It is expected to be particularly severe for
the high-mass cases C and E, which peak between the mass ranges of Type
II SN and PISNe.  As the right panel of Figure~\ref{imfcasefig} shows,
the metal-free star formation histories of a \tree{5}{11}{3}{8} tree are
not significantly different for the cases A, B, and D, with variations of
less than a factor of two at $z = 10 - 20$.  But for the more top-heavy
cases C and E the metal-free star formation rate is quite different,
and could be used as an IMF indicator. These cases may be excluded by
the r-process elements and [Zn/Fe] and produce a noticeably poor MDF,
but illustrates the key point that observed star formation histories in
the early universe can now be connected explicitly to Galactic chemical
evolution and may yield important information on the IMF.

Lucatello et al. (2005) followed another nucleosynthetic approach to
constraining the early IMF. They used the \feh\ $\approx -2$ onset of s-process
elements from AGB stars to constrain the IMF to have $\langle \sigma
\rangle = 1.18$ and $m_c = 0.79$ ($\langle M \rangle = 1.8$ \msun). When
used in this model for $Z < Z_{crit}$, their proposed IMF gives a Population
III number fraction \f0\ = 0.2, far above the observed constraint.
This discrepancy is probably related to the fact that their model has no
leverage on high-mass stars.  This nucleosynthesis model does have such
leverage. The permitted IMFs here, with $m_c \gtrsim 10$ and $\sigma
\sim 1.0$, easily satisfy the constraints imposed by their analysis of
s-process elements, and so confirm their suggestion of a more massive
primordial IMF, but skewed even further toward massive stars.

\subsection{Behavior of ``Minihalos''}
\label{minihalosection} 

The fiducial \tree{5}{11}{3}{8} tree contains approximately 1 million
total progenitor halos extending down to $T_{vir} = 10^3$ K. Halos
between $10^3$ K and $10^4$ K are generally termed ``minihalos'', and
are thought to cool only with molecular hydrogen, while more massive
halos can cool at least partly by emitting recombination emission of H,
which is ionized above $T \approx 10^4$ K.  Being small in mass ($M
\simeq 10^5 - 10^6$ \msun\ at $z \sim 10 - 30$), minihalos also may
be easily ionized by main-sequence massive stars, and even disrupted
completely by energetic supernovae. This model has, up to this point,
treated minihalos as ``closed boxes'' which do not experience ionization
or mechanical feedback. The strongest constraint on the primordial IMF,
\f0, does not depend on the assumed minimum virial temperature in the
tree (see Figure~\ref{covar-fig}), but the other constraints do depend
somewhat on the behavior of minihalos, and thus may be modified by
the feedback effects of massive stars.  This section develops some limiting
cases of radiative and mechanical feedback to assess the sensitivity of
the IMF constraints to these processes.  These feedback effects will
be more fully explored and connected to a Galactic chemical evolution
model in future work.

First I consider the local effects of ionization by massive stars. It is
possible for massive stars to produce \ion{H}{2} regions that completely
ionize the gas bound to minihalos. This process has been studied by
Kitayama et al. (2004) over a range of minihalo masses and assumptions
about stellar mass. They derived a critical halo mass below which
a minihalo is completely ionized for a given stellar mass (see their
Equation 14).  This feedback mechanism removes from the gas reservoir
the entire gas mass of halos that possess a massive star sufficient to
completely ionize its gas, according to the Kitayama et al. criterion.
This assumption represents an extreme, since it does not return this gas
to the reservoir as if it had recombined and cooled at a later time (this
time is estimated by Kitayama et al. at a few tens of millions of years).
Returning the recombined gas to the reservoir would allow subsequent
metal-enriched star formation (with metals ejected by the same massive
star that ionized it), so including feedback increases \f0\ as much as
possible by inhibiting star formation in small halos after the first,
metal-free generation. The overall increase in \f0\ is 20\% or less over
its value in a tree with no feedback.  The corresponding change in the
\f0\ constraint on the IMF is shown with the dashed contour in panel
A of Figure~\ref{imffig}, which marks the contour for \f0\ = 0.0019
including a 20\% increase in \f0\ to account for radiation feedback.
This effect is modest because it touches only a small number of halos
and a small fraction of the total gas budget.  Future work will need to
focus on more detailed treatments of this feedback mechanism to implement
cases in between the fiducial model and this extreme modification,
and to assess its overall importance to chemical evolution.

The other possible feedback mechanism which can affect chemical evolution
is the mechanical disruption of small halos by supernovae. PISNe
are particularly good at disrupting small halos ($M \lesssim 10^7$
M$_{\odot}$) as they release up to $\sim 10^{53}$ erg per event
(HW02). This feedback mechanism violates the assumption of a closed
box from which gas and metals do not escape.  Because this effect is
preferentially caused by PISNe at the higher mass end, it can invalidate
the limits placed above on [Zn/Fe] and [r/Fe], which implicitly assume
that the high iron yields of high-mass PISNe are retained in their parent
halos and dilute the Zn and r-process yields of core-collapse SNe.
Thus it is important to assess the potential effect of this feedback
mechanism on the limits of interest, the IMF constraints, and on Galactic
chemical evolution in general.

A simple prescription expresses mechanical feedback.  VMSs are allowed
to form as usual, but when one of them experiences a PISN, a check is
done to see whether the total energy of the supernova (seen in HW02
Figure 1) is larger than the binding energy of the gas in the parent
dark matter halo. If so, this gas reservoir is permanently removed from
the tree and assumed to reside in the diffuse IGM, where it is no longer
tracked. The effect of this feedback mechanism on \f0\ has already been
taken into account as ionization, since SNe that disrupt their parent
halos have already ionized them.  The more top-heavy IMFs disrupt up
to two third of all minihalos in the tree.  The main effect of this SNe
``blowout'' is to remove from the available reservoir of metals the very
high iron yields that dilute the products of core-collapse SNe and allow
the limits from detailed nucleosynthesis placed above in Section~5.3.
In the presence of feedback, these limits are easily achieved and the
upper end of the IMF is difficult to constrain because it leaves no
detailed nucleosynthetic signature.  If star formation begins in earnest
in halos with $T_{vir} \lesssim 10^4$ K, a wide range of weak-VMS IMFs
are formally permitted. However, even accounting for PISNe feedback,
these IMFs produce inferior fits to the MDF (see Figure~\ref{imfcasefig}).
Nevertheless, these effects may make it it may be difficult to uncover
evidence of the absence of VMSs from the IMF.

While SNe feedback complicates two possible nucleosynthetic limits
on the early IMF, it opens another potential avenue for detailed
constraints at the high-mass end.  By removing from the gas reservoir
material which is otherwise eligible for later star formation, and by
affecting potentially more than one-half of low-mass branches in the
tree, this mechanical feedback has a strong effect on the total budget
of ionizing photons that can escape and reionize the IGM. Preliminary
calculations suggest that for (50, 1.2), which produces 96\% of its
Fe with VMSs and so lies outside the [Zn/Fe] constraint from above,
loses 30\% of its cumulative budget of ionizing photons (up to $z =
10$) to halo disruption, thereby dropping from $\gamma_0 = 82600$ to an
effective value of 57800. This value still meets the fiducial constraint
of $\gamma_0 \geq 34000$. This reduction is less than expected given
that more than 2/3 of the minihalo branches in the tree terminate in a
PISN. Future models that treat reionization and chemical evolution in
detail will need to fully explore this feedback process.

\section{General Results from the Model} 

This section describes some general results derived from the chemical
evolution model that do not depend on or relate directly to the IMF, which
has been the focus of the paper. These results include information from
detailed chemical evolution histories of metal-poor stars, the end of the
first stars epoch.

\subsection{Results on Detailed Nucleosynthesis}

A given merger tree with chemical evolution allows for calculation of the
chemical histories of individual metal-poor stars, for the first time
in their proper context.  These histories include the number, masses,
metallicities, and supernova times of all progenitor SNe. This information
is critical to properly interpreting the relative element abundances in
metal-poor stars (Beers \& Christlieb 2005). There are two approaches
to using this information to understand the SNe yields of the first
stars. The first is to apply a ``basis set'' of theoretical yields to
the calculated distribution of progenitor masses and metallicities, and
then compare the calculated relative abundances to the data to adjust the
IMF parameters and optimize the fit. The second is to attempt to extract
intrinsic yields directly from the data empirically, with assistance from
a chemical evolution model. The former approach, which is commonly
used in studies of the metal-poor stars, is complicated by the fact
that there is really no suitable ``basis set'' of theoretical supernova
yields. Inspection of the yield tables in Woosley and Weaver (1995) shows
that theoretical yields vary dramatically with mass and metallicity,
and sparsely sample the available ranges in these variables. There are
also significant uncertainties associated with the choice of supernova
energy and mass cut between the compact remnant and the ejecta. Recent
models (e.g. Umeda \& Nomoto 2005) have introduced additional parameters
associated with mixing of material across the mass cut and a wider range
of explosion energies, in an effort to better match the observed Fe-peak
elements in Galactic EMPs. Finally, model Type II core-collapse SNe still
do not spontaneously explode - the SN energy must still be introduced
by hand. These factors taken together mean that attempting to assemble a
true ``basis set'' of model yields is quite difficult - no single group
or study has densely sampled the full parameter space needed to map
out the range of masses and metallicities probed here and in the data.
Also, this approach would introduce many more tunable parameters into
the model with sure loss of confidence in the results.

\begin{figure*}
\centerline{\epsfxsize=\hsize{\epsfbox{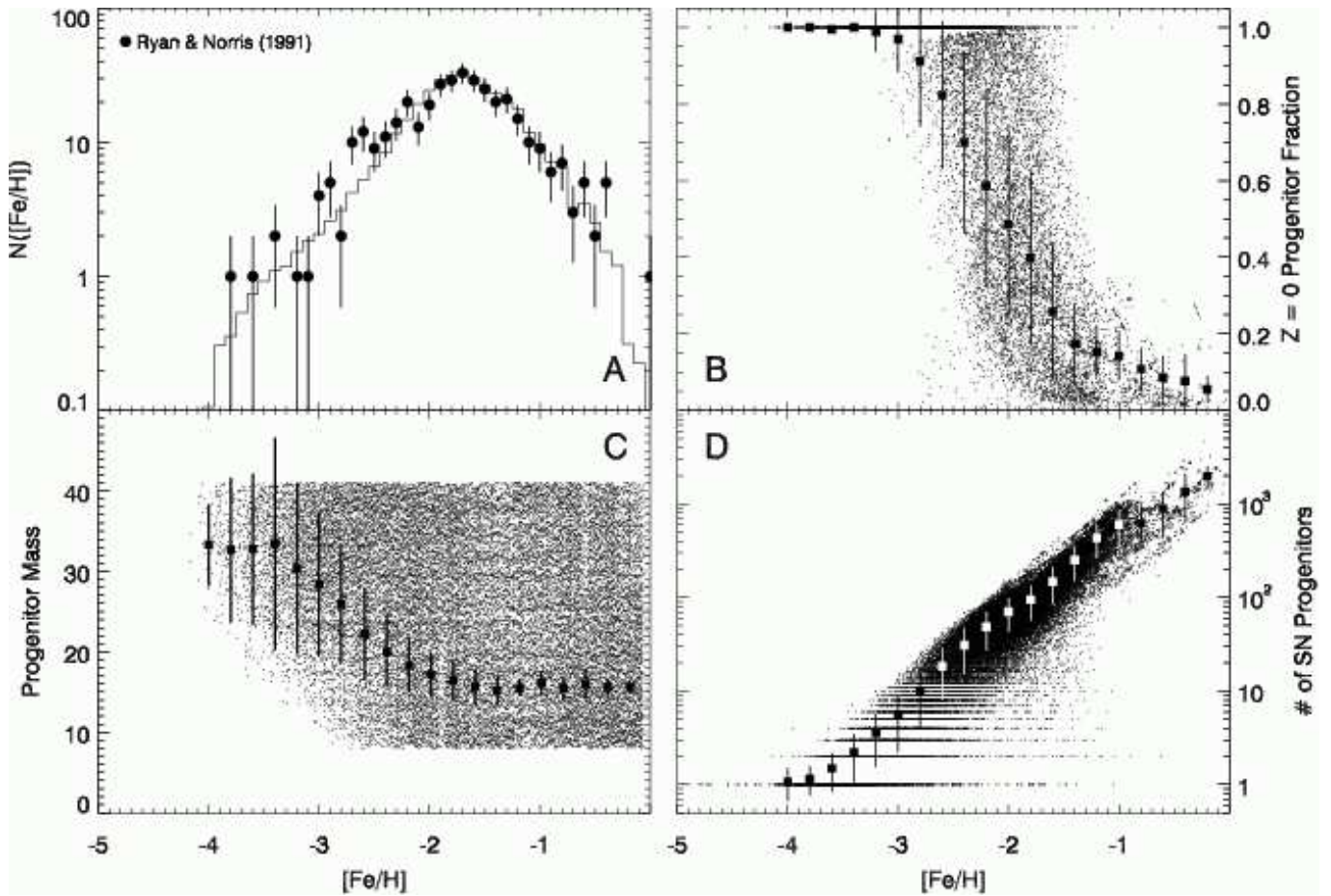}}} 
\figcaption{Detailed chemical evolution histories for model metal-poor
stars.  Panel A: MDF calculated for a single \tree{5}{11}{4}{8}
tree, IMF case A, and $Z_{crit} = -4$.  Panel B: As a function of
final stellar metallicity, the fraction of its SN progenitors that
were metal-free. Squares mark the mean in 0.2 dex bins, with 1 $\sigma$
scatter shown by error bars.  Between \feh\ = -3 and \feh\ = -2, this
fraction drops dramatically from 1 to $ < 10$ \%. Panel C: As a function
of final metallicity, the distribution of progenitor masses for each
metal-poor star. Squares show the mean progenitor mass in 0.2 dex bins,
while the error bars show the scatter in the mean mass.  Lower mass SN
progenitors enter later, with the 8 - 12 \msun\ stars that probably give
rise to the r-process elements entering at \feh\ $\sim -3$. Panel D: As a
function of final stellar metallicity, the number of SN progenitors. For
a given \feh, the number of progenitors can show a wide scatter. The mean
number of progenitors is 10 at roughly \feh\ $= -2.8$.  \label{progfig}}
\end{figure*}

For these reasons I adopt the latter approach, namely, to attempt to
extract information, and perhaps detailed yields, directly from the
data with assistance from the chemical evolution model. This approach 
complements the direct modeling of observed abundance patterns 
by potentially revealing new trends that result from the underlying 
processes of chemical evolution.  This method is enabled by the
detailed chemical evolution histories available from the model presented
here. Figure~\ref{progfig} shows detailed results for the IMF
test-case A in a \tree{5}{11}{4}{8} tree, and shows the power of
these models to interpret the metal-poor stellar abundance data in the
proper context. Panel A shows the overall MDF for this fiducial tree.
Panel B shows, for all the metal-poor stars in the tree, the fraction
of their progenitors that were metal-free stars. Panel C shows the mass
distribution of progenitor SN for each metal-poor star, and panel D shows
the number of progenitors for each star. In the latter three panels,
the mean quantity in 0.2 dex bins is plotted with a square, and the
1$\sigma$ scatter in that quantity is plotted with error bars.

A number of important results appear in this figure. First, the model
is constructed to preserve a great amount of useful information about
the history of chemical enrichment in the model Galaxy, most importantly
the mass and metallicity distribution of SNe precursors for metal-poor
stars for a given parameter set and tree realization.  Second, combining
the results of panels B and D, I find that below \feh\ $\sim -3$,
EMP stars have 1 - 10 SNe precursors, all of which are metal-free.
In fact, the average \feh\ $\sim -4$ (UMP) star had exactly one
metal-free precursor. This removes the dependence of the observed
abundances on precursor metallicity and leaves only precursor mass as
a variable. Thus the variation of [Zn/Fe] with \feh\ below \feh\ $= -3$
(Cayrel et al. 2004) may be explained simply as an effect of SN precursor
mass, with more massive, short-lived precursors, $M \simeq 30 - 40$ \msun,
coming in first at \feh\ $\lesssim -4$, followed by longer-lived precursors
of 8 -- 10 \msun\ only at \feh\ $\sim -3$. Put another way, the mean mass
of precursors declines from $\sim 35$ to $\sim 15$ \msun\ over the range
\feh\ = $-4$ to $=2$. Finally, the predicted number of precursors per
metal-poor star, and their mass and metallicity distribution, can be used
to assess the scatter in observed [X/Fe] ratios.  This conclusion has
interesting implications for the interpretation of observed r-process
abundances that will be explored in a future paper.  

These key results represent a major step toward the final goal
of extracting intrinsic yields of the first SNe directly from the
data. This study has developed the framework needed to perform this
analysis within the proper context of galaxy formation. Future work
will focus on exploring the full range of possibilities in parameters,
on understanding the systematic errors, and on assessing the volume of
data and level of data quality needed to give satisfactory results.

\subsection{End of the First Stars Epoch}

There are no known locations in the local universe where metal-free gas
is known to exist.  Intensive efforts to discover galaxies at $z > 6$
(the end of reionization) have not yet turned up any galaxies that
are clearly metal-free, or even known to have \feh\ $\lesssim -2$,
the lowest metallicity seen in the local universe. Thus metal-free star
formation must have been curtailed at some point in the early universe,
probably in the first 1 Gyr after the Big Bang. In the absence of any
observational information about metal enrichment of galaxies before $z
= 6$, theorists have approached the problem by trying to estimate the
time needed for galaxies to enrich themselves, or neighboring galaxies,
using simulations and semi-analytic models in a similar spirit to the
one presented here.

If the metal enrichment history of the Galaxy's halo is typical, star
formation is already substantially enriched with metals even by $z =
20$. According to Figure~\ref{imfcasefig},  for IMFs A, B, and D, only 10
- 20\% of star formation is still metal-free at $z = 20$, where the total
rate of truly metal-free star formation in the Galaxy's history peaks
at $0.2$ \msun\ yr$^{-1}$. By $z \sim 6$, metal-free star formation has
declined to 3\% of the total star formation rate in halo progenitors. This
metal-free star formation is isolated to small ``leaf'' halos which have
just virialized, not distributed throughout halos that are also forming
metal-enriched stars. If the Milky Way halo is a representative sample
of high-redshift star formation, there is a quite low probability of
finding a truly metal-free star-forming galaxy at $z \sim 6$, 
owing to the self-enrichment of halos by local star formation. 

\begin{figure}
\centerline{\epsfxsize=\hsize{\epsfbox{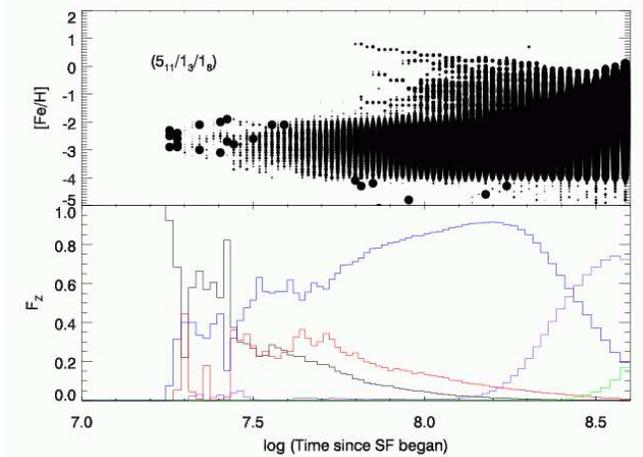}}} 
\figcaption{Upper panel: The distribution of stellar metallicities and birth times
for the fiducial \tree{5}{11}{3}{8} tree and IMF case A, showing the
spread in times for a single [Fe/H], or the spread in [Fe/H] at a
given time. The symbol sizes are proportional to the number
of stars in each event. Clearly time and metallicity are not perfectly
correlated. For instance, stars with \feh\ = -3 form over a period of 500
Myr.  Lower panel: The fraction of stars in different metallicity ranges,
as a function of time since the onset of star formation. Stars with \feh\
$<-4$ in black, $-4\leq$ \feh\ $\leq-3$ in red, $-3 \leq$ \feh\ $\leq -2$
in blue, $-2 \leq$ \feh $\leq -1$ in purple, and $\leq -1$ \feh\ $\leq 0$
in green. Star formation is essentially all metal free for $\sim 10^7$
yr, and down to $\sim 20$\% of the total by 30 Myr. \label{epochfig}}
\end{figure}

Figure~\ref{epochfig}, which shows star formation times and metallicities
for a \tree{5}{11}{3}{8} tree, showing the spread in times for a single
[Fe/H], or the spread in [Fe/H] at a given time. Clearly time and
metallicity are not perfectly correlated and low-metallicity stars form
throughout the history of Galactic halo formation. In the lower panel, the
curves show the fraction of star formation in different metallicity ranges
as a function of time since the onset of star formation in the tree. For
the first 10 Myr, star formation is exclusively metal-free (black curve),
but by 30 Myr it has become only 20\% of all star formation.

By relating high-$z$ star formation rates directly to Galactic chemical
evolution, this model can provide additional unique constraints on the
nature of the earliest stellar populations. Future work will focus on
expanding the chemical evolution model applied here to the Galaxy to a
population of galaxies at an arbitrary redshift in population syntheses
that can be compared directly to high-redshift data and to the Galactic
halo stars.

\section{Discussion}

By connecting chemical evolution to Galaxy formation, this model
has placed meaningful constraints on the mass function of the first
stars using a broad range of observational information. I now consider
how these constraints might improve with additional data, and how theory
might guide future experiments.

The samples of metal-poor Galactic halo stars are expected to
grow dramatically in the next decade owing to systematic surveys
dedicated to finding and studying large stellar populations in
the outer galaxy.  The SEGUE extension to the Sloan Digital Sky
Survey\footnote{http://www.sdss.org} will discover up to 20000 stars
with \feh\ $< -2.0$, 2000 with \feh\ $< -3.0$, 200 with \feh\ $< -4.0$,
and a few stars with \feh\ $ \sim -5$ to $-6$, should they exist. SEGUE
will construct the sample and estimate stellar parameters and \feh. These
stars can then be studied at high resolution and good S/N with 8 m class
telescopes to read from them the mass assembly and chemical evolution
history of the Galaxy.  Another such survey is the WFMOS project,
which is planned to obtain high-resolution spectra of 1 million stars
with a 1000-fiber optical spectrograph mounted on the Subaru telescope
\footnote{http://www.gemini.edu/science/aspen/wfmos-sciobj.pdf}.
This survey is also directed at the earliest phases of chemical
evolution in the galaxy. With $> 10^5$ stars these surveys could move
the \f0\ constraint near or past the \f0\ $\sim 10^{-4}$ contour in
Figure~\ref{imffig}, or even find a true Population III star. Their real
power may lie in their ability to provide enough data for individual
SNe yields to be extracted empirically.  If one of these surveys does
manage to discover a true metal-free star, it will be very interesting
in its own right.  If no truly metal-free old stars are found in these
expanded samples, they will tighten the constraint on \f0\ and force
the first-stars IMF to even more skewed high masses.

By obtaining colors, radial velocities, and metallicities for large
samples of Galactic halo stars, these surveys hope to achieve ``chemical
tagging'', or the identification of distinct groups of kinematically and
chemically related stars that would arise from a single coherent formation
history. This result would constitute a major achievement in ``near-field
cosmology'' (Freeman \& Bland-Hawthorn 2002).  Even if they succeed, these
ambitious efforts will boost, and not moot, theoretical efforts
like the one reported here.  These surveys can provide additional known
and reliable laboratories to study detailed chemical evolution, where
we now operate on sample of indistinguishable stars without regard to
their individual kinematic histories and no reliable way to group them
except by metallicity.  At the same time this observational technique
will provide critical information on the CDM history of the Galaxy,
such that a detailed actual history may some day be constructed with
a combination of theory and data.  There will still be ample room for
theory to reconstruct the possible chemical histories and relate them to
other indicators, and connect them to the ``cosmology'' in ``near-field
cosmology'', with a new level of rigor and accuracy. As stated before,
this theoretical study is only the beginning step in this long process.

The real power of this chemical evolution model to constrain reionization
will come when it is applied to a wide range of halo masses and redshifts,
so that various reionization histories and chemical evolution trajectories
can be related in detail for full galaxy populations. This effort will
be reported in a later paper.

My results also point to an important lesson about the power of
theoretical calculations of supernova yields to constrain the mass
function of the first stars. As shown in \S~5, the yields of PISNe are
difficult to completely rule out based on comparisons of theoretical
yields directly to the data.  This difficulty is represented here by
the constraints offered by [Zn/Fe] and the difficulty of pushing it
any further. By comparison with the robust constraint on the strong
VMS hypothesis offered by the r-process elements, the argument based on
[Zn/Fe] seems crude. This point illustrates the importance of identifying
clean signatures for particular mass ranges of the IMF. If, for instance,
the r-process is isolated to the 8 - 12 \msun\ for low-metallicity stars,
the constraints on the upper end of the IMF can be improved. Where
possible, theorists should attempt to isolate such clear indicators of
stellar mass in their increasingly sophisticated models.

\section{Conclusions}

I have presented a new treatment of chemical evolution in the earliest
phases of Galaxy formation. This model is motivated by and integrated
with the theory of hierarchical galaxy formation. As a result, this model
can be formally connected to high-redshift star formation, allowing
detailed connections between disparate phenomena.  This simple model
can reproduce the main properties of the Galactic halo metallicity
distribution as well or better than classical theories of chemical
evolution (see Oey 2002 and Qian and Wasserburg 2002 for recent
examples), which cannot be formally connected to high-redshift data in
such detail. The basic structure of the model follows chemical evolution
of stars and star-forming gas within dark matter halos whose masses are
specified by the hierarchical theory of structure formation and which
merge and evolve according to a well-known framework.  Within each
halo, the treatment of chemical evolution is more akin to standard
theory, but with some improvements, mainly related to the stochastic
treatment of early star formation. The model advances the interpretation 
of metal-poor stars in the Galaxy by generating detailed chemical
evolution histories for each metal-poor star in the model.  Most
importantly, this model allows {\it quantitative} hypothesis testing of
novel theoretical ideas that have been lacking in material connection to
the available data. This new framework is simple, flexible, and fast, so
that almost any new idea about early stellar and interstellar evolution
can be readily incorporated and connected to the whole of the present
and future data.

Based on the hundreds of individual models run with this new framework,
and the best available data, I draw the following key conclusions:
\begin{itemize}
\item[1.] The non-detection of true Population III stars in the Galactic halo
      strongly suggests that there is some critical metallicity,
      $Z_{crit} \lesssim 10^{-4}$, below which low-mass star formation
      was inhibited.  The precise value of $Z_{crit}$ cannot be determined
      within the range $0 < Z_{crit} \lesssim -4$ because the Galactic MDF
      has not yet shown a statistically significant deficit of low-\feh\
      stars (except for true Population III).
\item[2.] The frequency of true Population III stars, \f0, is robust indicator
      of the IMF below $Z_{crit}$ because variations in model parameters
      for halo formation and mixing are overwhelmed by changes in the IMF.
      The IMFs that best fit all the available data from metal-poor stars
      and reionization have log-normal parameters $m_c = 6 - 35$ \msun\
      and $\sigma = 0.3 - 1.2$, and mean masses $\langle M \rangle =
      8 - 42$ \msun. These masses are close to the predictions from
      theoretical models of primordial star formation that include
      formation feedback.
\item[3.] For a given set of parameters, individual chemical evolution
      histories can be calculated for each model metal-poor star.
      Population II stars with \feh\ $\leq -3$ (EMPs and below) have 1 - 10
      metal-free SN progenitors, suggesting that the trends in EMP ratios
      such as [Zn/Fe] result from variations in the masses of precursor
      SNe. This result is a significant advance in comparing detailed yields of
      the first SNe to theoretical calculations.
\end{itemize}

With the advent of new large telescopes devoted to high-redshift studies
of galaxies, and large surveys of metal-poor stars in the Milky Way and
other galaxies, we are entering a new age of rich information about early
star formation and chemical enrichment.  This study is the first step
in a large effort to use the tools of theory to relate different types
of observation together.  As these issues develop, the new theoretical
framework developed here will be extended and improved to analyze new
information in unified, coherent way.  It may turn out that the history
of cosmic star formation is written in the Galaxy. Only by combining
all of the state-of-the-art data can we be sure of confidently reaching
detailed conclusions about the first stars in the universe.

\acknowledgements
This work would not have been possible without the generous support of the
Department of Astronomy and Astrophysics at the University of Chicago. It
is my pleasure to acknowledge Jim Truran, Joss Bland-Hawthorn, and
Vikram Dwarkadas for interesting and helpful conservations, and Tim
Beers for these and detailed comments on the manuscript. This study
grew out of ongoing work with my valued collaborators Aparna Venkatesan
and Mike Shull, and I am grateful for their continuing stimulation and
encouragement, and for comments on the manuscript.

\appendix

This appendix discusses the many details involved in the halo merger
tree model. Most of the extended Press-Schechter formalism relies on the
expression for the variance of the initial Gaussian random density field,
expressed in Fourier $k$-space:

\begin{equation}
S(M) \equiv \sigma^2(M) = \frac{1}{2\pi^2} \int^{\infty}_{0} k^2 P(k) \bar{W}^2(kR) dk
\end{equation}
In theory the integral is over the full range $k = 0 - \infty$, but in
practice the range is considered to be $k = 0.001 - 1000$ Mpc$^{-1}$. In
this relation, $P(k)$ is the matter power spectrum:
\begin{equation}
P(k) = \xi k^n T^2(k).
\end{equation}
Assuming $n=1$ and the transfer function $T(k)$ derived by Bardeen et
al. (1986),
\begin{equation}
T(k) = \frac{\ln (1+2.34q)}{2.34q}[1+3.89q+(16.1q)^2+(5.46q)^3+(6.71q)^4]^{-1/4},
\end{equation} where $q = k/(\Omega_M h^2 {\rm Mpc}^{-1})$. In Equation
(A1), the Fourier transform of the top-hat filtering function is
given by:
\begin{equation}
\bar{W} = \frac{3}{(kR)^3}[\sin (kR) - (kR)  \cos (kR)]
\end{equation}
Finally, the normalization constant $\xi$ is found by evaluating the
integral (A1) and equating the result to $\sigma _8^2$, the variance
over spheres of $R = 8 h^{-1}$ Mpc. 
The linear growth function from
Carroll, Press, \& Turner (1992) extrapolates the linear density 
field to the present day: 
\begin{equation}
D(z) = \frac{g(z)}{g(0)(1+z)}
\end{equation} where
\begin{equation}
g(z) = \frac{5}{2}\Omega_M \left[ \Omega _M^{4/7} - \Omega _{\Lambda} +
     \left(1+\frac{\Omega _M}{2}\right) \left( 1+\frac{\Omega _\Lambda}{70}\right) \right]^{-1}
\end{equation}
With all these pieces in place, the Press-Schechter function
describing the number density of dark matter halos of between mass $M$
and $M +dM$ is:
\begin{equation}
ndM = \sqrt{\frac{2}{\pi}} \frac{\bar{\rho}}{M} \frac{\delta _c
      (z)}{\sigma ^2 (M)} {\rm exp} \left[- \frac{\delta ^2 _c (z )}{2 \sigma ^2
      (M)}\right] dM
\end{equation} In this expression, $\delta_c (z)$ is the
redshift-dependent critical overdensity for collapse, $\delta_c (z) =
\delta _{c,0}/D(z)$ and $\delta _{c,0} = 1.686$.

\renewcommand{\arraystretch}{1.2}
\begin{deluxetable}{ccl}%[!hb]
\tablecolumns{3} \tablewidth{0pc} \tablecaption{Model Constraints} 
\tablehead{Datum & Value & Constraint }
\startdata
Galactic Halo MDF & shape & disk/halo separation, IMF \\ 
Population III fraction, \f0\              & $\leq 0.0019$ & Lower end of IMF \\ 
Ionizing efficiency $\gamma_0$   &  $\geq$ 34000  & Lower end of IMF \\ 
Maximum [Zn/Fe], 8 - 40 \msun\   & $\leq$ 1.0 & Upper end of IMF \\ 
$F_{VMS}$, VMS iron budget        & $\leq$ 0.5 & Upper end of IMF \\ 
Fraction of r-process enriched stars & $\geq$ 0.82 & Upper end of IMF 
\enddata
\label{constraint-table}
\end{deluxetable}


\begin{references}

\reference{abel} Abel, T., Bryan, G. L., \& Norman, M. L. 2002, Science, 295, 93 
\reference{AL96} Adams, F. C.,  \& Laughlin, G. 1996, \apj, 468, 586
\reference{ag} Aguirre, A., Hernquist, L., Schaye, J., Katz, N., Weinberg, D. H., \& Gardner, J. 2001, \apj, 561, 521
\reference{pop2} Baade, W. 1944, \apj, 100, 137  
\reference{bard} Bardeen, J. M., Bond, J. R., Kaiser, N., \& Szalay, A. S. 1986, \apj, 304, 15
\reference{heres} Barklem, P., et al. 2005, A\&A, in press 
\reference{BPS} Beers, T. C., Preston, G. W., \& Schechtman, S. A. 1992, \aj, 103, 1987
\reference{tcb05} Beers, T. C. 2005 private communication
\reference{beers} Beers, T. C., \& Christlieb, N., 2005, ARA\&A, in press 
\reference{BCL} Bromm, V., Coppi, P., \& Larson, R. B. 2001, \mnras, 328, 969  
\reference{BL02} Bromm, V., \& Loeb, A. 2002, NewA, 9, 353     
\reference{BL03} Bromm, V., \& Loeb, A. 2004, Nature, 425, 812
\reference{Brommrev} Bromm, V. \& Larson, R. B. 2004, ARA\&A, 42, 79 
\reference{burr00} Burris, D., et al. 2000, \apj, 544, 203
\reference{Carney96} Carney, B. W., Laird, J. B., Latham, D. W., \& Aguilar, L. A. 1996, \aj,
                      112, 668
\reference{c02} Carretta, E., Gratton, R., Cohen, J.G., Beers, T. C.,
        \& Christlieb, N. 2002, \aj, 124, 481
\reference{cpt} Carroll, S. M., Press, W. H., \& Turner, E. L. 1992, ARA\&A, 30, 499
\reference{Cen03b}Cen, R. 2003, \apj, 591, L5
\reference{cfw} Ciardi, B., Ferrara, A., \& White, S. D. M. 2003, MNRAS, 344, 7
\reference{HE0107} Christlieb, N., et al. 2003, Nature, 419, 904
\reference{cayrel} Cayrel, R., et al. 2004, A\&A, 416, 1117
\reference{cohen} Cohen, J. G., et al. 2004, \apj, 612, 1107
\reference{ftc} Fields, B. D., Truran, J. W., \& Cowan, J. J. 2002, \apj, 575, 845
\reference{figer} Figer, D. 2005, Nature, 434, 192 
\reference{Frebel} Frebel, A., et al. 2005, Nature, 434, 871
\reference{jbh} Freeman, K. C., \& Bland-Hawthorn, J. 2002, ARA\&A, 40, 487 
\reference{gnedin} Gnedin, N. 1998, \mnras, 294, 407
\reference{HH} Haiman, Z., \& Loeb, A. 1997, \apj, 483, 21 
\reference{hh03} Haiman, Z., \& Holder, G. P. 2003, 595, 1
\reference{hw02} Heger, A., \& Woosley, S. 2002, \apj, 567, 532 (HW02) 
\reference{ky} Kitayama, T., Yoshida, N., Susa, H., \& Umemura, M. 2004, \apj, 613, 631 
\reference{LC93} Lacey, C., \& Cole, S. 1993, \mnras, 262, 627 (LC93) 
\reference{l73} Larson, R. B. 1973, \mnras, 161, 133
\reference{s99} Leitherer et al., 1999, \apjs, 123, 3 
\reference{ll00} Liddle, A. R., \& Lyth, D. H. 2000, Cosmological Inflation 
    and Large-Scale Structure (Cambridge: Cambridge Univ. Press) 
\reference{bl01} Loeb, A., \& Barkana, R. 2001, ARA\&A, 39, 19
\reference{Luc05} Lucatello, S., Gratton, R. G., Beers, T. C., Carretta, E. 2005, \apj, in press
\reference{mfr} Madau, P., Ferrara, A., \& Rees, M. J. 2001, \apj, 555, 92
\reference{mr02} Malhotra, S., Rhoads, J. E. 2002, \apj, 565, L71 
\reference{mcw95} McWilliam, A., Preston, G. W., Sneden, C., \& Searle,
        L. 1995, \aj, 109, 2757
\reference{mw02} Mo, H. J., \& White, S. D. M. 2002, \mnras, 336, 112
\reference{mso00} Oey, M. S. 2000, \apj, 542, L25
\reference{mso02} Oey, M. S. 2002, \mnras, 339, 849
\reference{OP} Omukai, K., \& Palla, F. 2002, ApJ, 589, 677
\reference{omukai} Omukai, K., Tsuribe, T., Schneider, R., \& Ferrara, A. 2005, \apj, 626, 627 
\reference{PS} Press, W. H., \& Schechter, P. 1974, \apj, 187, 425
\reference{QW02} Qian, Y.-Z., \& Wasserburg, G. J. 2002, \apj, 567, 515 
\reference{rn91} Ryan, S. G., \& Norris, J. E. 1991, \aj, 101, 1865 (RN91)
\reference{IMF} Salpeter, E. E. 1955, \apj, 121, 161
\reference{ss05} Santoro, F., \& Shull, J. M. 2005, \apj, in preparation
\reference{ssf} Scannapieco, E., Schneider, R., \& Ferrara, A. 2003, \apj, 589, 35 
\reference{sch} Schaerer, D. 2002, A\&A, 382, 28
\reference{sch03} Schaye, J., Aguirre, A., Kim, T. S., Theuns, T., Rauch, M.,
                    \& Sargent, W. L. W. 2003, \apj, 596, 768
\reference{sch} Schneider, R., Ferrara, A., Natarajan, P., \& Omukai, K. 2002, \apj, 571, 30 
\reference{sk99} Somerville, R. S., \& Kolatt, T. S. 1999, \mnras, 305, 1 (SK99)
\reference{sp99} Somerville, R. S., \& Primack, J. R. 1999, \mnras, 310, 1087
\reference{wmap} Spergel, D. N., et al. 2003, \apjs, 148, 175
\reference{spitzer} Spitzer, L. 1978, Physical Processes in the Interstellar Medium, (New York: Wiley) 
\reference{Tm02} Tan, J., \& McKee, C. M. 2002, in ``The Emergence of Cosmic Structure'',
                Proceedings of the AIP, 666, 93
\reference{Tm02} Tan, J., \& McKee, C. M. 2004, \apj, 603, 683 
\reference{jwt02} Truran, J. W., Cowan, J. J., Pilachowski, C. A., \&
      Sneden, C. 2002, \pasp, 114, 1293
\reference{ts00} Tumlinson, J., \& Shull, J. M. 2000, \apj, 528, L65
\reference{tgs01} Tumlinson, J., Giroux, M. L., \& Shull, J. M. 2001, \apj, 550, L1 
\reference{tvs04} Tumlinson, J., Venkatesan, A., \& Shull, J. M. 2004, \apj, 612, 602 (TVS04)
\reference{tsv03} Tumlinson, J., Shull, J. M., \& Venkatesan, A. 2003, \apj, 584, 608 (TSV03)
\reference{un04} Umeda, H., \& Nomoto, K. 2005, \apj, 619, 427 
\reference{vts03} Venkatesan, A., Tumlinson, J., \& Shull, J. M. 2003, \apj, 584, 621
\reference{W94} Woosley, S. E., et al. 1994, \apj, 433, 229
\reference{ww95} Woosley, S. E., \& Weaver, T. A. 1995, \apjs, 101, 181
\end{references}
\end{document}